\documentclass[%
 reprint,
 amsmath,amssymb,
 aps,
pra,
]{revtex4-1}

\usepackage{graphicx}
\usepackage{dcolumn}
\usepackage{bm}
\usepackage{braket}
\usepackage{amsmath}

\begin{document}

\preprint{APS/123-QED}

\title{Chip-scale Simulations in a Quantum-correlated Synthetic Space}

\author{Usman A. Javid$^{1}$}
\author{Raymond Lopez-Rios$^{1}$}%
\author{Jingwei Ling$^{2}$}%
\author{Austin Graf$^{1}$}%
\author{Jeremy Staffa$^{1}$}%
\author{Qiang Lin$^{1,2}$}%
\email{qiang.lin@rochester.edu}
\affiliation{$^{1}$Institute Of Optics, University of Rochester, Rochester, NY 14627, USA}
\affiliation{$^{2}$Department of Electrical and Computer Engineering, University of Rochester, Rochester, NY 14627, USA}
\date{\today}

\begin{abstract}
An efficient simulator for quantum systems is one of the original goals for the efforts to develop a quantum computer \cite{aspuru2012photonic}. In recent years, synthetic dimension in photonics \cite{synthetic} have emerged as a potentially powerful approach for simulation that is free from the constraint of geometric dimensionality. Here we demonstrate a quantum-correlated synthetic crystal, based upon a coherently-controlled broadband quantum frequency comb produced in a chip-scale dynamically modulated lithium niobate microresonator. The time-frequency entanglement inherent with the comb modes significantly extends the dimensionality of the synthetic space, creating a massive nearly $400\times400$ synthetic lattice with electrically-controlled tunability. With such a system, we are able to utilize the evolution of quantum correlations between entangled photons to perform a series of simulations, demonstrating quantum random walks, Bloch oscillations, and multi-level Rabi oscillations in the time and frequency correlation space. The device combines the simplicity of monolithic nanophotonic architecture, high dimensionality of a quantum-correlated synthetic space, and on-chip coherent control, which opens up an avenue towards chip-scale implementation of large-scale analog quantum simulation and computation \cite{analog-sim,aspuru2012photonic,AA} in the time-frequency domain. 

\end{abstract}

\maketitle


\begin{figure*}
\centering
  \includegraphics[scale=0.47]{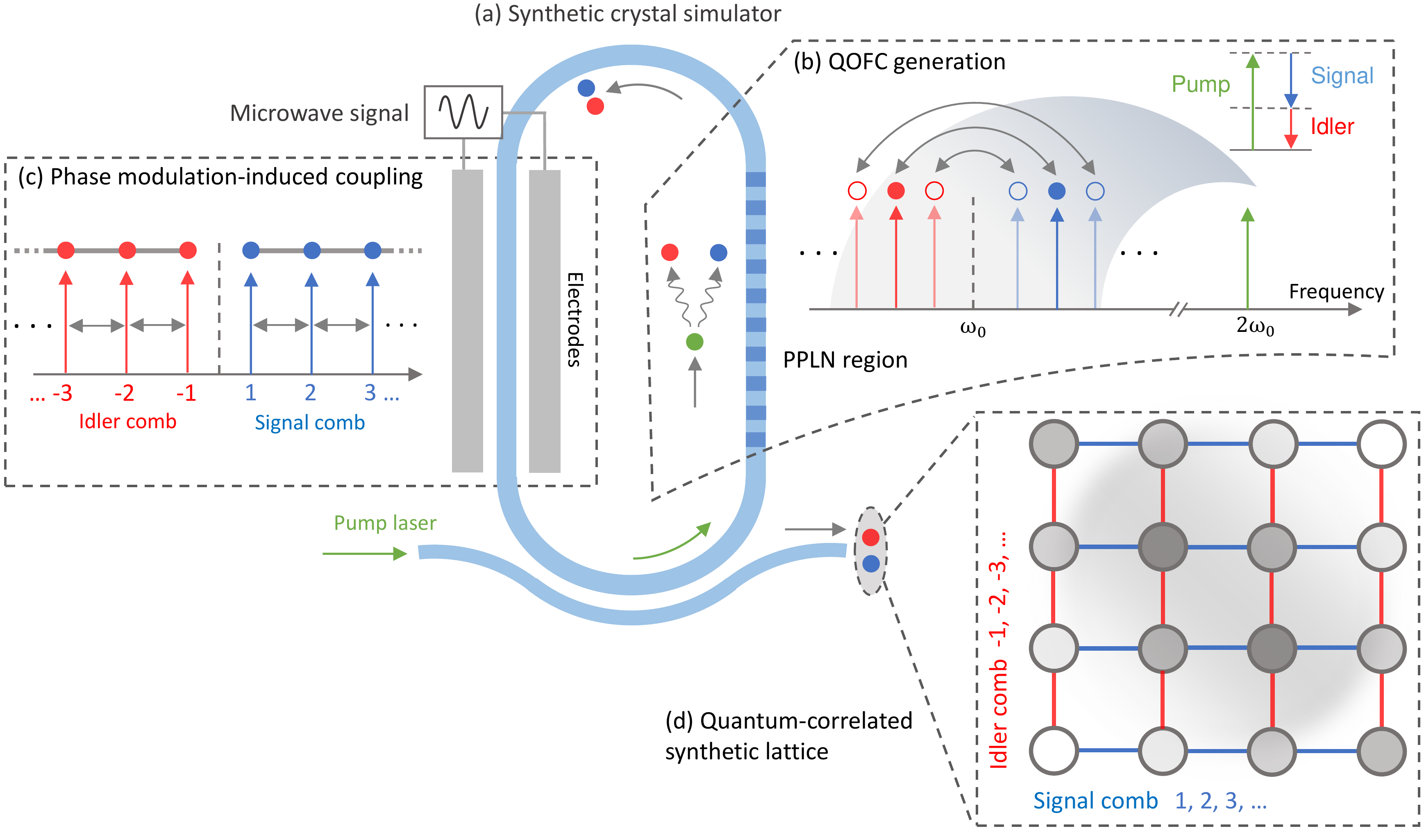}
  \caption{Concept of a quantum-correlated synthetic crystal. (a) A nanophotonic lithium niobate racetrack resonator. The synthetic lattice is constructed in two steps. First a periodically-poled region in the resonator generates pairs of time-frequency entangled photons within a frequency comb creating the nodes of the synthetic lattice as shown in (b). Then an elctro-optic modulator directly embedded inside the resonator creates coupling within the comb lines forming a tight-binding lattice for each photon, as shown in (c). Combining these two effects creates a two-dimensional quantum-correlated synthetic lattice as illustrated in (d).}
  \label{fig1}
\end{figure*}

\section{\label{sec:level1}Introduction}

Despite decades of effort, a fully-configurable and error-corrected optical quantum computer remains out of reach. The non-interacting nature of photons, weak nonlinearities, and unavoidable losses prevent efficient implementation of logic gates and protocols for fault-tolerant computation. However, it is still possible to achieve a quantum advantage in simulation using non-interacting particles, even in the presence of noise and loss \cite{AA,AA-loss}. Such intermediate-scale machines are designed specifically to simulate one physical system, sacrificing universality much like an analog computer. To that end, linear optical circuits have been widely implemented \cite{circuit1, circuit2, circuit3, circuit4, circuit5, circuit6, circuit7, circuit8, aspuru2012photonic, wang2020integrated}, where the position or path information of photons is employed for the  simulation. This approach, however, faces challenges in scaling up for simulating complex problems which require ever-increasing physical space, sometimes needing hundreds of elements with both optical and electrical interconnects \cite{large1,large2, wang2020integrated}, that would impose impractical requirement on the fabrication and its precision of the underlying photonic integrated circuits. 

\par An alternate approach harnesses a so-called \emph{synthetic dimension}. The frequency space is a good example of this \cite{synthetic}. Here the simulation runs on photons moving between distinct frequency modes, all of which can occupy the same physical space, for example, a spatial mode of a cavity. Increasing the dimensions of the system requires minimal increase in the complexity of the architecture. Furthermore, the frequency domain provides a complete equivalent to linear optical quantum computing (LOQC) \cite{freq-loqc}, with all operations available through frequency mixing interactions. Investigations on these synthetic spaces have attracted significant interest recently, with great potential for simulating a variety of condensed-matter phenomena and topological effects \cite{syn-theory1, syn-theory2, syn-theory3, syn-theory4, syn-theory5, syn-theory6, bartlett2021deterministic}. So far, the experimental implementations have been limited to the classical regime, where light from a laser populates the synthetic frequency lattice \cite{syn-exp1, syn-exp2, syn-exp3, syn-exp4, syn-exp5, syn-exp6, syn-exp7, syn-exp8, syn-exp9, syn-exp10, syn-chip, syn-chip2}. Simulations based on non-classical light, however, would provide unique insights into transport phenomena at the quantum scale \cite{quantum1, quantum5} and bring advantages offered by quantum mechanics that are inaccessible to classical simulation spaces \cite{phase-retr, quantum2, quantum3, quantum4, bartlett2021deterministic}. The primary challenge lies in producing a synthetic space that is capable of generating quantum states of light while simultaneously being able to coherently control their evolution depending on the specific simulation problem. In particular, realization of such a synthetic space on a chip-scale platform would offer tremendous benefits in resource efficiency, system scalability, and operation stability \cite{wang2020integrated} that are challenging for table-top systems  \cite{syn-exp1, syn-exp2, syn-exp3, syn-exp4, syn-exp5, syn-exp6, syn-exp7, syn-exp8, syn-exp9, syn-exp10}.  

Here we demonstrate an on-chip quantum-correlated synthetic crystal, which is characterized by non-classical correlations between lattice sites. Such a crystal significantly extends the dimensionality of the synthetic space via quantum correlations inherent with energy-time entangled photons, in contrast to the sole frequency degree of freedom available in the classical regime \cite{syn-exp1, syn-exp2, syn-exp3, syn-exp4, syn-exp5, syn-exp6, syn-exp7, syn-exp8, syn-exp9, syn-exp10, syn-chip, syn-chip2}. We implement this concept using a coherently-controlled quantum optical frequency comb (QOFC) composed of $\sim$800 single-photon comb modes, produced inside a dynamically modulated thin-film lithium niobate microresonator. The time-frequency entanglement associated with the spontaneous parametric down-conversion (SPDC) process introduces long-range quantum correlations between the signal and idler frequency combs, while an electro-optic modulator implemented within the QOFC generation process produces nearest-neighbour coupling, eventually creating a tight-binding quantum correlated synthetic crystal (Fig.~\ref{fig1}). 

With such a system, we are able to utilize the evolution of the generated quantum correlations to perform a series of simulations in the \emph{time and frequency correlation space} created by the entangled photons. 
We demonstrate two-particle quantum random walks simulated on the \emph{bi-photon spectral correlation space}. We further simulate Bloch oscillations of an electron in a crystal lattice in the presence of an external gauge potential, on the \emph{bi-photon temporal correlation space}. Finally, we drive the lattice sites into the strong coupling regime and observe multi-level Rabi oscillations where adjacent lattice sites exchange energy faster than the lifetime of photons inside the resonator. This series of proof-of-principle simulations clearly demonstrate the capability and versatility of the device for simulating physics of crystal lattices as well as individual atoms. The device presented here creates a massive nearly 400$\times$400 mode quantum correlated synthetic space with electrically-controlled tunability, opening up a path towards running complex large-scale simulations and computations for near-term analog quantum simulators.

\begin{figure*}
\centering
  \includegraphics[scale=0.95]{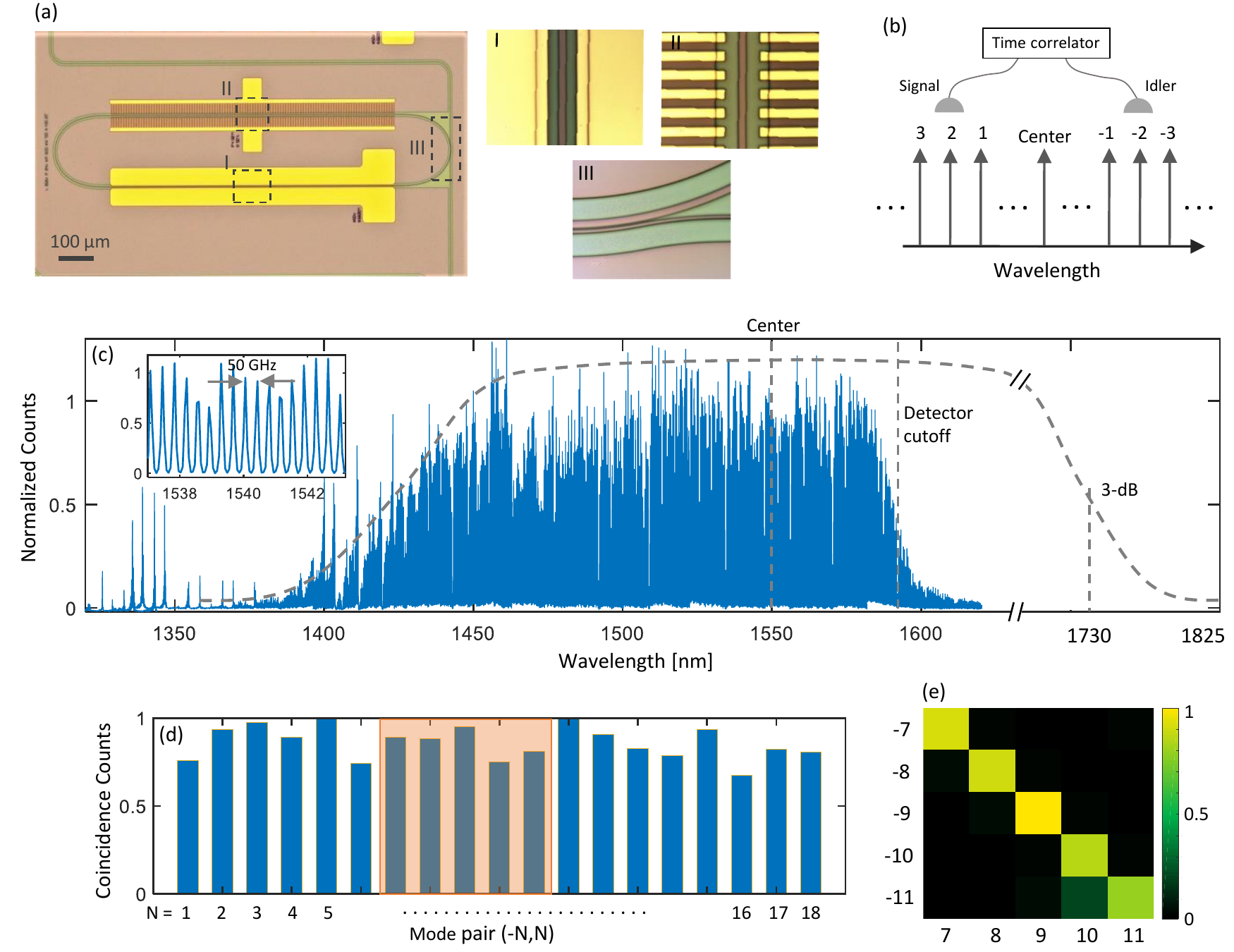}
  \caption{Characterization of the bi-photon quantum optical frequency comb (QOFC). (a) Microscope image of a fabricated device, where I shows a section of the embedded electro-optic modulator, II shows a section of the periodically poled waveguide region for QOFC generation, and III shows a section of the pulley coupling waveguide for broadband external coupling. (b) Schematic showing the correlation measurement where the spectral correlation of the photon pairs is performed by scanning a pair of single-photon detectors through the comb modes and counting the coincidence events with a time correlator. (c) Recorded spectrum of produced QOFC, showing discrete comb modes lighting up with a 50 GHz mode spacing (inset). The spectrometer cuts-off at 1590~nm, preventing the spectral measurement at longer wavelengths. We estimate that the comb spectrum reaches $\sim$1800~nm on the longer wavelength side as indicated by the dashed line. (d) Coincidence histogram for different signal-idler mode pairs, each of which has signal and idler frequencies equally spaced from the the center of the generated spectrum. Strong coincidences are measured due to energy conservation of signal-idler photon pair in the SPDC process. This presents as a bright diagonal in a two-dimensional joint spectral intensity (JSI) plot a shown in (e). When the detectors are aligned to off-diagonal mode pairs, there are no measurable coincidences as expected from a pure SPDC process.}
  \label{fig3}
\end{figure*}

\begin{figure*}
\centering
  \includegraphics[scale=1]{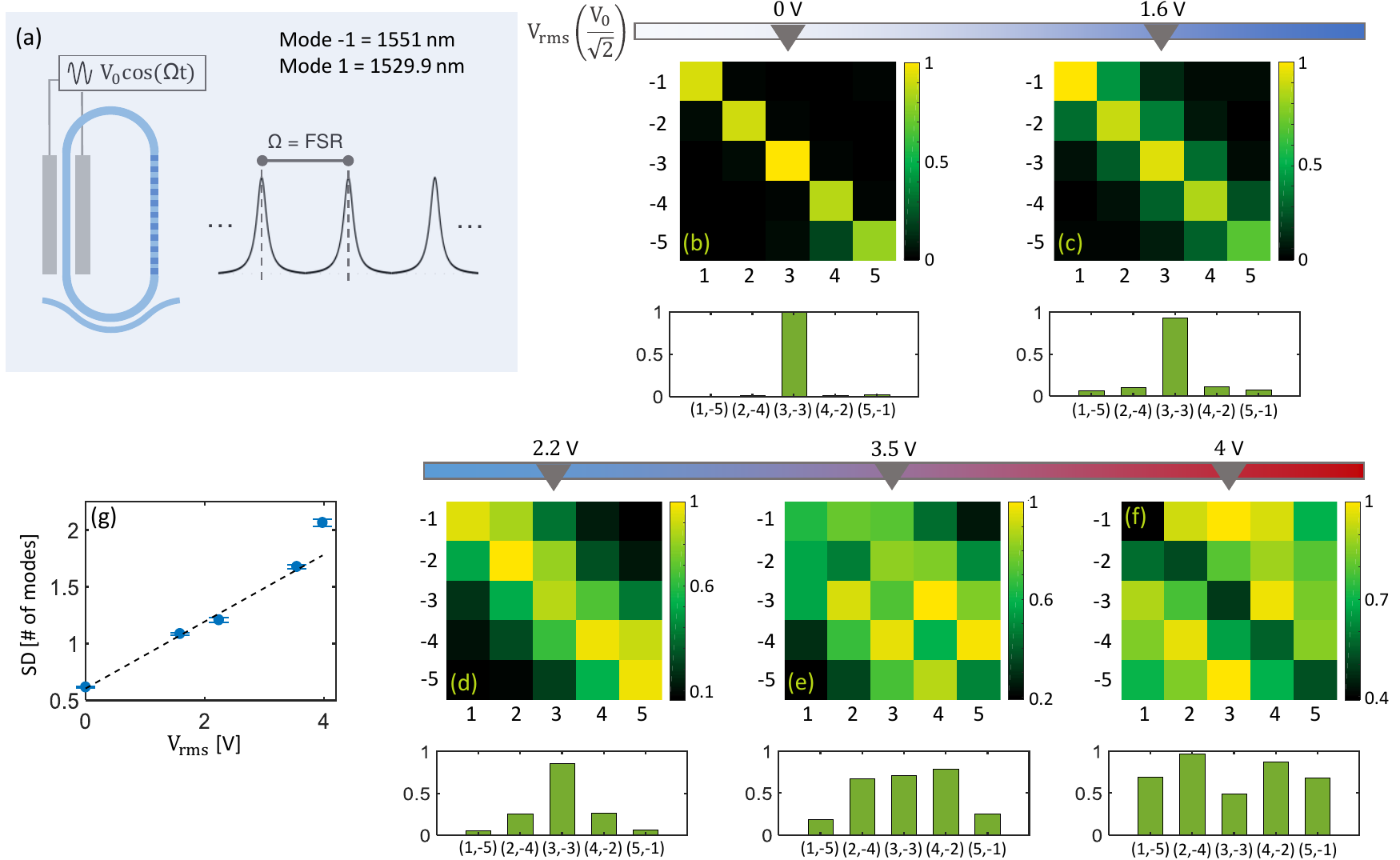}
  \caption{Quantum random walk of correlated photons in the synthetic lattice. (a) The microwave signal frequency $\Omega$ is tuned to match the FSR of the resonator at 50 GHz, and its amplitude V$_0$ is varied. (b)-(f) Measured JSI of the photon pairs in a 5$\times$5 mode space for increasing microwave signal amplitude with the signal modes on the x-axis and the idler modes on the y-axis. The bar plots below the JSI show the values along the secondary diagonal of the matrices indicating the spread of the JSI. The standard deviation (SD) of the random walk for each modulation amplitude is plotted in (g) along with a linear fit. Error bars indicate one standard deviation of uncertainty.}
  \label{fig4}
\end{figure*}

\section{Results}
\subsection{Device design and the synthetic space}

Figure~\ref{fig1} shows the schematic of the device concept. The synthetic lattice is formed by the frequency modes of an optical cavity containing simultaneously a QOFC generator and a high-speed phase modulator. Experimentally, we implement this system on a nanophotonic lithium niobate chip with a racetrack resonator as shown in Fig.~\ref{fig1}(a). In order to be able to run simulations in this space, there are three design requirements. First, the entire synthetic space has to be able to be populated with quantum light. This is done by dispersion engineering the resonator geometry for broadband SPDC \cite{dispersion}. A section of the resonator is periodically-poled to match the refractive index gap between the pump frequency and the frequency of the generated photons. This creates a bi-photon frequency comb that forms the skeleton for our synthetic space as shown in Fig~\ref{fig1}(b). Second, we establish a tight-binding crystal by implementing nearest-neighbor coupling with frequency modulation. For this we place a pair of electrodes designed to operate at microwave frequencies as shown in Fig~\ref{fig1}(c). When the material refractive index is modulated at a frequency that matches the cavity free spectral range (FSR), light can scatter to adjacent modes using sum and difference frequency interactions creating a tight-binding system, with a coupling strength determined by the modulation depth (see Supplementary Information (SI) for a detailed theoretical treatment). This, along with a pair of entangled photons, creates a two-dimensional synthetic space for running simulations with the dimensions created by the chain of resonances formed by each photon of the pair as shown in Fig~\ref{fig1}(d). This space is unique since it uses both classical correlation from frequency modulation and quantum correlation from SPDC to create a lattice and is therefore capable of investigating classical and quantum simulation phenomena. Finally, we must be able to extract light from each mode within the synthetic space with equal weight. This is implemented by using a coupling waveguide with an optimized wrap-around geometry placed aside with the resonator (see SI for details).

\par With this design, the device is fabricated on a thin-film lithium niobate chip. The SI document provides details on device fabrication, poling and characterization of its linear properties. When the device is pumped at a resonance at 776~nm, we obtain a broadband QOFC in the telecom band, with a 3-dB half-width of 19.5~THz centered at 1552~nm as shown in Fig.~\ref{fig3}(c) and a 50-GHz mode spacing as shown in the inset. Due to the spectrometer's cutoff wavelength at 1590~nm, we only have access to the signal (blue) side of the spectrum. However, the energy conservation constraint of the SPDC process implies that the spectral extent of the comb should be symmetric around its center wavelength, which infers a 3-dB spectral bandwidth of 39~THz that contains approximately 780 modes of the resonator. This provides a massive higher-dimensional space to construct a synthetic crystal for simulations or computational experiments, utilizing coherent control of the comb lines implemented directly within the generation process which has been inaccessible in other quantum frequency comb platforms \cite{comb-review}. We further characterize the frequency correlations for a small 18-mode subspace by filtering pairs of modes that are symmetric around the spectral center and recording coincidence counts between the mode pair. As shown in Fig.~\ref{fig3}(d), all the mode pairs exhibits fairly uniform correlation amplitudes, indicating an equal probability of producing biphotons across the QOFC spectrum. Moreover, We also measure these correlations for pairs of modes that are not frequency matched in a 5-by-5 matrix, obtaining a joint-spectral intensity (JSI) map as shown in Fig.~\ref{fig3}(e). The results presented in Fig.~\ref{fig3}(e) clearly show the well-known anti-correlation of the photon-pair modes expected from SPDC, indicating strong frequency correlations and a clean photon-pair generation process. The details on the experimental setup are provided in SI.

\begin{figure*}
\centering
  \includegraphics[scale=1]{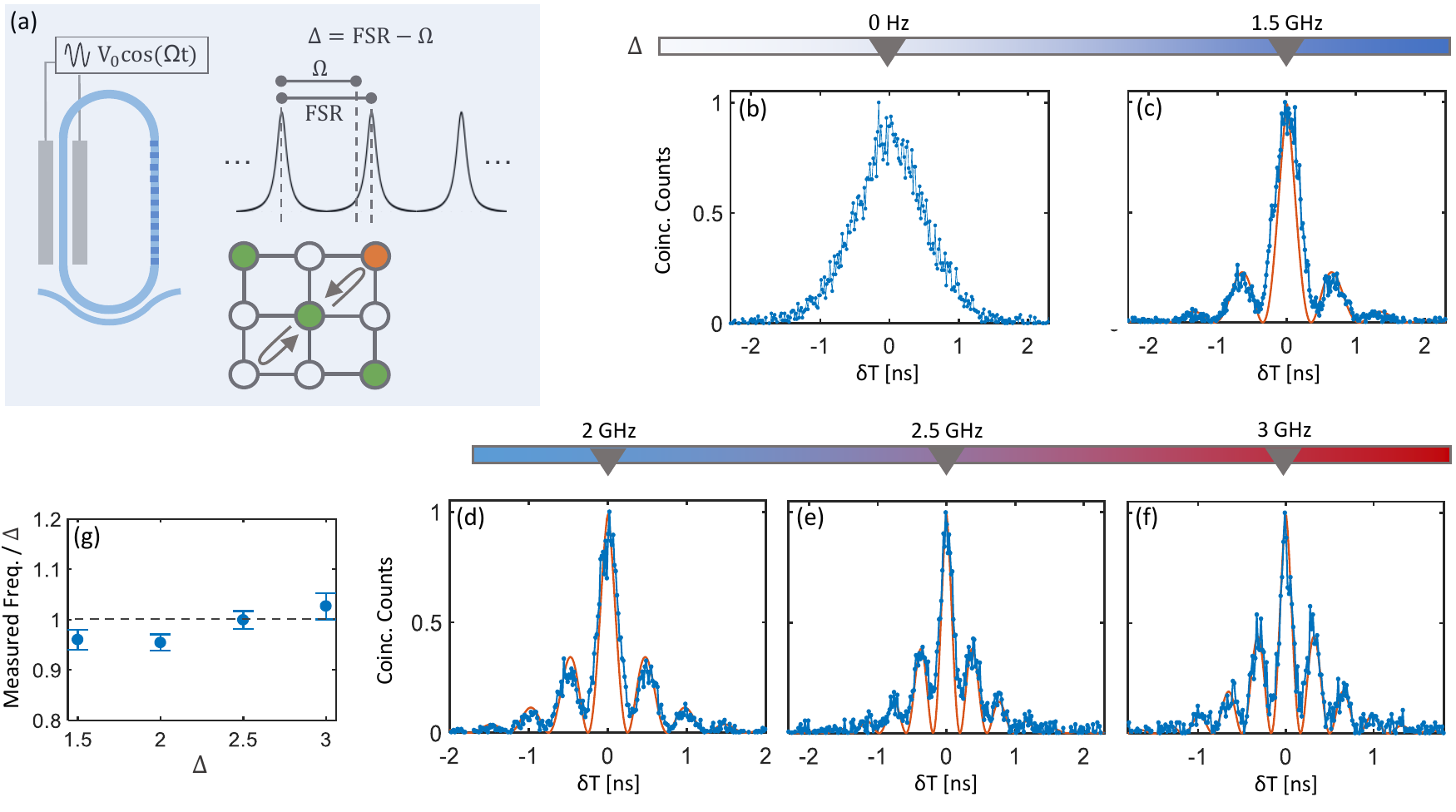}
  \caption{Simulating Bloch oscillations on the bi-photon temporal correlations. (a) The modulation signal is detuned from the resonator FSR by an amount $\Delta$ while the modulation amplitude V$_0$ is kept constant. A lattice site (indicated in orange) one position away from the main SPDC diagonal (indicated in green) is picked to measure the bi-photon temporal correlations for varying values of $\Delta$. Arrows indicate the oscillatory movements of the bi-photon correlations with the detuning. (b)-(f) Bi-photon temporal correlations obtained by accumulating a histogram of coincidences for differences in arrival time ($\delta$T) of the two photons for increasing values of $\Delta$. Orange line plots indicate theoretical fitting of the data. (g) The ratio of the measured oscillation frequency to the detuning obtained by a sinusoidal fit of the coincidence histogram after correcting for the exponential decay envelope. Error bars indicate a 95\% confidence bound on the frequency fit.}
  \label{fig5}
\end{figure*}

 \subsection{Simulations in the time-frequency correlation space}
 \subsubsection{Quantum random walks of correlated photons}
 With the strong quantum correlations of the produced QOFC modes, we are in a position to create a tight-binding synthetic crystal. This is realized by turning a microwave drive on and matching its frequency to the resonator FSR at 50~GHz. The first experiment is a quantum random walk of a particle in a tight-binding crystal. This is a natural outcome of this system since at any given mode, a photon has an equal probability of scattering to either a lower frequency or a higher frequency mode, with an amplitude that can be controlled with the microwave power. This creates a continuous-time random walk for the two photons in a 1-D chain of modes. However, the bi-photon frequency correlation is a 2-D space as shown in the JSI plot in Fig~\ref{fig3}(e). We will track the trajectory of the JSI during the random walk by measuring coincidences at each pair of modes in a 5$\times$5 space with increasing modulation amplitude. Although we make measurements in a small sub-space of the total available bandwidth due to time constraints, the rest of the synthetic lattice will behave in the same way. Figure~\ref{fig4} shows the results of this measurement. As we increase the modulation amplitude, the bi-photon correlation starts to spread perpendicular to the anti-diagonal, and the spread increases with increasing microwave signal power. From the initial state in Fig~\ref{fig4}(b), we can observe photons jumping multiple modes within their lifetime before escaping the resonator, sometimes up to four modes at strong modulation powers. The random walk of the photon pairs tends to reduce their frequency correlation, moving them from a strongly-correlated state towards a separable state. This is quite understandable as the walk randomizes the correlation information that the photons initially contain when generated. This is also inline with the theoretical treatment of this system, presented in the SI. Figure~\ref{fig4}(g) plots the standard deviation (SD) of the JSI as a function of the modulation voltage. The SD is scaled such that an unmodulated SPDC process gives a zero spread (see Methods for details). The data fits well with a line as expected from a quantum random walk \cite{var1,var2}. The slight deviation from linearity at the highest modulation amplitude is due to the onset of nonlinearity in the gain of the microwave amplifier used in the experiment as well as modulation on the pump laser (see SI for a note on this discrepancy). It is important to note that this evolution of the spectral correlations is realized within the SPDC generation process without any post-processing, in contrast to other similar random walk experiments that have to rely on complex circuit structures such as a network of beam splitters or waveguides \cite{random-waveguide} or post-processing tools such as pulse shapers and filters \cite{var2}. This experiment provides a much simpler architecture -- an optical cavity with a time-dependent length, for these experiments. Moreover, the fully integrated device approach indicates the possibility for a tunable photon pair source with active control over the time-frequency entanglement.

 \begin{figure*}
\centering
  \includegraphics[scale=1]{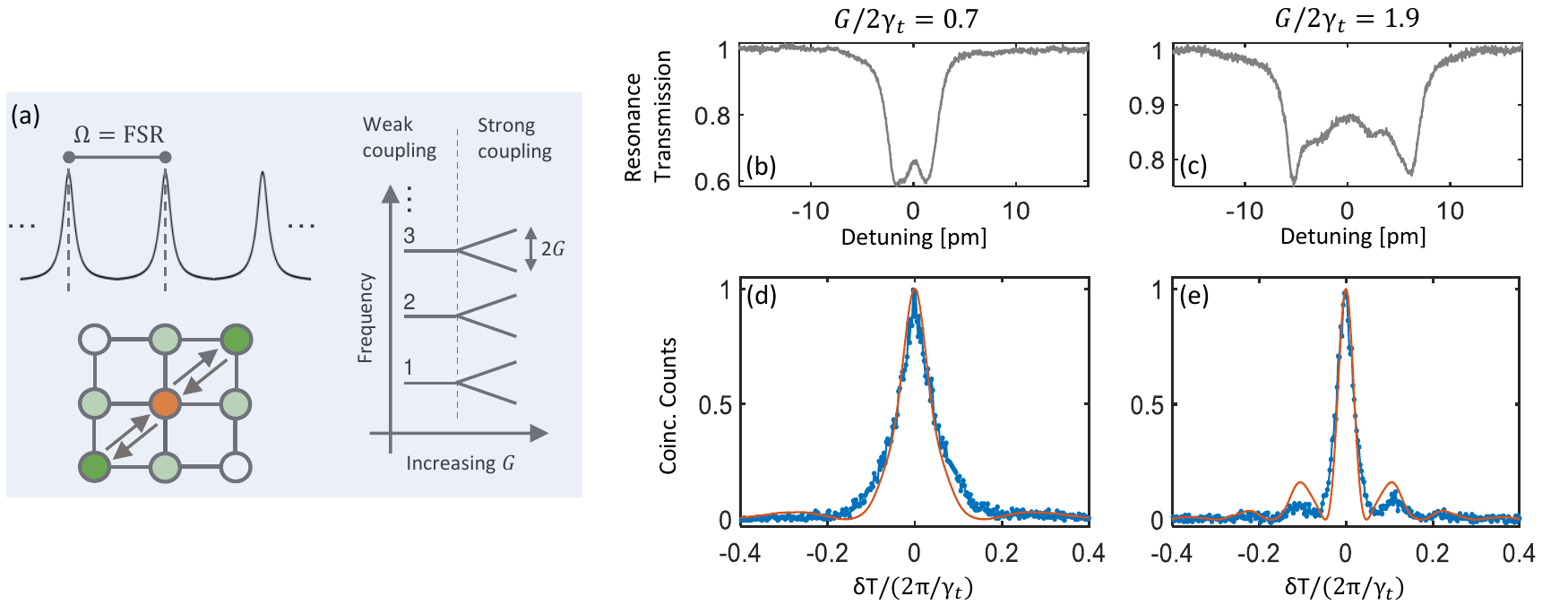}
  \caption{Strong coupling and Rabi oscillations. (a) The modulation signal frequency $\Omega$ is tuned again to the resonator FSR. The modulation signal amplitude V$_0$, represented here as the effective mode coupling strength $G$, is varied. When the coupling strength exceeds the resonator linewidth, the resonator modes split into two dressed modes spaced from each other by 2$G$, indicating the onset of strong coupling regime. We pick a site corresponding to a signal-idler mode pair on the main SPDC diagonal indicated in orange for the bi-photon temporal correlation measurement. Arrows indicate exchange of photons between adjacent lattice sites. Transmission spectrum of a signal-mode cold-cavity resonance is plotted for two values of the coupling strength indicating (b) weak coupling and (c) strong coupling regime with $\gamma_t$ being the loaded resonator linewidth. The corresponding bi-photon coincidence histogram for the chosen lattice site are plotted in (d) and (e). Here we observe Rabi oscillations in the strong coupling regime. Orange lines indicate theoretical results, and the time axis is normalized to the photon lifetime.}
  \label{fig6}
\end{figure*}

\subsubsection{Synthetic Electric Fields and Bloch Oscillations}
 \par Apart from the modulation amplitude, the high-speed phase modulator embedded in the microresonator enables free tuning of the modulation frequency $\Omega$. Physically, a detuning $\Delta = \text{FSR}-\Omega$ imparts a phase on the coupling in the tight-binding model (see SI for details), which is directly equivalent to the presence of an external gauge potential applied on a solid-state system \cite{gauge}. For a constant detuning, this corresponds to the simulation of a constant electric field acting on electrons in the crystal lattice, which leads to Bloch oscillations. By introducing this detuning, we can simulate an effective electric field for the photon pairs, forcing them into oscillations. Simulations of Block oscillations have been realized in many experiments in the past (see for example \cite{bloch1,bloch2}). However, our system creates a unique situation since all the modes in the synthetic space are equally likely to be populated with a photon, and all the modes exchange photons between them at the same rate. This means that on average, photon flux in each mode has no time dependence, and no net movement of photons can be observed with intensity measurements. The temporal dynamics of the oscillation are revealed only in the second-order intensity correlation of the QOFC modes. To observe such Bloch oscillations, we pick an off-diagonal term in the frequency space which, ordinarily, will not frequency match SPDC, as shown in Fig~\ref{fig5}(a). Any coincidences in this mode pair will only occur when the bi-photon frequency correlations spread due to the random walk. For this simulation, we collect a histogram of coincidence counts for differences in arrival times between the two photons. This gives the bi-photon temporal correlation function, which typically shows an exponentially decaying envelope corresponding to the photon lifetime of the loaded resonator. At zero detuning, that is all we see as shown in Fig~\ref{fig5}(b). When a detuning is introduced, oscillations emerge with a period matching the detuning frequency as shown in the coincidence histograms in Fig~\ref{fig5}(c)-(f), indicating an oscillatory probability of detecting a photo pair in the selected mode pair. Initially, the modulation signal causes the bi-photon correlations to spread, but at a characteristic time corresponding to $2\pi/\Delta$, the effective electric field forces the photons to backtrack their movement, returning to the original correlated spectrum. This process repeats until the photons escape the resonator and this movement is imprinted onto the coincidence histogram. Therefore, the simulation of Bloch oscillations is run in the temporal correlation space, where the effective electric field forces the photon pairs to oscillate between a strongly correlated and a uncorrelated state in the frequency space. Fig~\ref{fig5}(g) plots the ratio of the oscillation frequency measured by a sinusoidal fit of the coincidence histogram to the microwave signal detuning (see methods for details), which shows good agreement. A theoretical treatment for this oscillatory evolution of the bi-photon correlations is presented in the SI.

 \subsubsection{Strong Coupling Regime and Rabi Oscillations}

 In systems of coupled modes, when the rate at which the modes exchange energy exceeds the system's decay rate, Rabi oscillations emerge. Typically seen in coherently driven atomic two-level systems \cite{Rabi}, Rabi oscillations are characterized by spitting of the energy levels into dressed states. The synthetic lattice can simulate these effects when the coupling strength of the signal and idler modes exceeds their linewidth. In this case, the lattice sites exchange energy several times before the photons escape the resonator. These oscillations, like Bloch oscillations, are imprinted in the bi-photon temporal correlations (at a zero detuning of the microwave frequency, $\Delta = 0$, while with a strong enough driving amplitude). To show this phenomenon, we utilized a different device with a cavity linewidth smaller than the one used in the previous two experiments. Figure~\ref{fig6} shows the results for a pair of comb modes on the central correlation diagonal. When the microwave driving amplitude is small, the system falls within the weak coupling regime as evident by the transmission spectrum of a cold cavity resonance shown in Fig.~\ref{fig6}(b). Accordingly, the recorded coincidence histogram of the mode pairs exhibits a normal exponentially decaying envelope as shown in Fig.~\ref{fig6}(d). However, with a strong enough microwave driving, we achieved a coupling strength nearly twice the linewidth of the cavity, as evident by the large mode splitting of the cold cavity resonance shown in Fig.~\ref{fig6}(c). Consequently, the recorded coincidence histogram becomes oscillatory in this strong coupling regime, as shown in Fig.~\ref{fig6}(e). The measured correlations are in good agreement with the theoretical expectations. 
 The Rabi oscillations observed here are unique given that we are dealing with a multi-level system instead of the traditional two-level system where these effects are typically studied. Here, the generated photons oscillate between three adjacent coupled modes. We also obtain a mode splitting that is close to twice the coupling strength, double that of a two-level system. SI provides a detailed theoretical treatment for the strong coupling regime. 
 
 Overall, the three experiments characterized on the bi-photon spectral and temporal correlations demonstrate the versatility and richness of the device physics available to us with the active control in the time-frequency domain.

\section{Discussion}
The ability to scale a quantum system to include enough particles and dimensions to the point where the computation becomes advantageous over a classical computer has become a critical metric for competing architectures. This is not only to prove engineering prowess of one platform over another, but also to settle a central conflict in computer science, complexity theory and quantum mechanics, i.e. that certain computational tasks scale faster on a machine using quantum states than what complexity theory suggests, and that such a machine can actually be realized \cite{shor}. To that end, several demonstrations have been made to varying degrees of success \cite{circuit1,supremacy1,supremacy2}. In this regard, this work demonstrates several useful properties of the monolithic LN nanophotonic platform and the fabricated device. First, the device design requires only a single microresonator with a time varying optical path length to create the synthetic space. Together with dispersion engineering, we are able to create a nearly 800-dimensional coupled-mode network. Secondly, the resonator is capable of generating quantum correlated photons in the entire space, requiring only a single-mode laser and a microwave signal as inputs to generate and coherently-control a QOFC, requiring no post-processing optics. This allows us to run both classical and quantum simulations on this device. For instance, the quantum random walk of correlated photons has features that are not present in random walks with classical states (see SI for a note). On the other hand, Bloch oscillations, an inherently classical simulation, are simulated on a quantum correlation function. The simplicity in the device design is also beneficial for scaling up this proof-of-principal demonstration to run complex computation. For instance, we have shown that the microwave signal can provide flexible control over the nature of the synthetic space. Taking this further, a multi-tone excitation can create short and long-range coupling in the lattice, and amplitude/phase modulation can be utilized to explore more complex physics. Furthermore, the highly multi-dimensional nature of the SPDC process in the lattice can be used for multi-photon generation, approaching the continuous variable regime, where the squeezing parameter is reasonably large. In this case, the correlation imparted by SPDC can be complemented by the correlation due to the resonator mode coupling to create cluster states for one-way computation \cite{continuous-cluster}. Our device is well-suited for this task.

\par To conclude, we have demonstrated on-chip generation and coherent control of an ultra-broadband bi-photon QOFC on the lithium niobate integrated photonic platform. We have utilized this QOFC to create a quantum-correlated synthetic lattice with nearly a $400\times400$ mode space. We have demonstrated the capability of this system by running three simulations: a two-particle quantum random walk, generation of a synthetic electric field and Bloch oscillations, and multi-level Rabi oscillations, on the bi-photon spectral and temporal correlation functions. The device presented here combines the simplicity of monolithic nanophotonic structure, a high-dimensional quantum-correlated synthetic space, and electrically-controlled tunability, which now opens up an avenue towards running complex large-scale simulations and computational tasks. We envision this work will motivate further investigations into chip-scale implementations of analog quantum simulation and computation in the time-frequency domain.

\section{Methods}
The QOFC spectrum in Fig.~\ref{fig3}(c) was obtained directly using an infrared spectrometer. The spectrometer camera captures a 70-nm bandwidth in a single exposure. The full spectrum covering 1300-1600~nm was pieced together by rotating the diffraction grating to change the center wavelength by 60 nm each time and capturing a spectrum with a one minute integration time. The detector of the spectrometer has a cutoff wavelength of 1590~nm. The spectral correlation measurements were made by scanning two tunable filters over the resonator modes. For the random walk data in Fig.~\ref{fig4}, each point on the JSI was taken for a integration time varying from 2 to 5 minutes depending on the modulation depth. This is because at high modulation, the generation efficiency is much lower due to the pump scattering away into unwanted frequencies which don't phase-match SPDC, thus requiring longer integration time. The JSI value at each data point in the matrices in Fig.~\ref{fig4} was obtained by integrating within the envelope of the coincidence histogram for the corresponding mode-pair. The Bloch oscillation plots were obtained for integration times varying from 30 minutes to an hour for the highest frequency oscillations. For details on the full experimental setup see SI. 
\par The variance of the random walk can be calculated as
\begin{align}
    \text{SD}^{2}=\frac{\sum_{X,Y} d^{2}_{X.Y} \text{JSI}_{X,Y}}{\sum_{X,Y}\text{JSI}_{X,Y}},
\end{align}
where $d_{X.Y}$ is the diagonal distance of each point on the JSI from the main matrix diagonal in integer units. This is used to evaluate spread of the random walk in Fig.~\ref{fig4}(g). The Bloch oscillation frequencies are measured by first correcting the coincidence histogram for the exponential decay envelope and then numerically fitting the flattened oscillations with a sinusoidal function. The frequency of the fitting function is varied until the RMS error of the fit reaches a minimum. the resulting frequency is plotted in Fig~\ref{fig5}(g) with the error bars representing a 95\% confidence bound on the frequency parameter.

\begin{acknowledgments}
The authors thank Olivier Pfister (University of Virginia) for useful discussions. This work is supported in part by the National Science Foundation (NSF) (OMA-2138174, ECCS-1810169, ECCS-1842691), the Defense Threat Reduction Agency-Joint Science and Technology Office for Chemical and Biological Defense (grant No. HDTRA11810047), and the Defense Advanced Research Projects Agency (DARPA) LUMOS program under Agreement No. HR001-20-2-0044. This work was performed in part at the Cornell NanoScale Facility, a member of the National Nanotechnology Coordinated Infrastructure (NNCI), which is supported by the National Science Foundation (Grant NNCI-2025233).

The project or effort depicted was or is sponsored by the Department of the Defense, Defense Threat Reduction Agency. The content of the information does not necessarily reflect the position or the policy of the federal government, and no official endorsement should be inferred.
\end{acknowledgments}

\section{Additional information}
\noindent\sffamily\textbf{Author contributions}:
Q. L. and U. A. J. conceived the experiment.  U. A. J., and J. S. conducted theoretical analysis and ran simulations for device design. U. A. J. and J. L. fabricated the device. U. A. J., R. L. R. and A. G. set up and ran the experiment. U. A. J. conducted data analysis and wrote the manuscript with contributions from all authors. Q. L. supervised the project.\\\\
\noindent\sffamily\textbf{Competing interests}
The authors declare no competing interests.\\\\
\noindent\sffamily\textbf{Data Availability}
Data is available on request. Correspondences should be sent to Q.L. (qiang.lin@rochester.edu).

\bibliographystyle{naturemag}
\bibliography{apssamp}

\end{document}


\preprint{APS/123-QED}

\title{Supplementary Information: Chip-scale Simulations in a Quantum-correlated Synthetic Space}

\author{Usman A. Javid$^{1}$}
\author{Raymond Lopez-Rios$^{1}$}%
\author{Jingwei Ling$^{2}$}%
\author{Austin Graf$^{1}$}%
\author{Jeremy Staffa$^{1}$}%
\author{Qiang Lin$^{1,2}$}%
\email{qiang.lin@rochester.edu}

\affiliation{$^{1}$Institute Of Optics, University of Rochester, Rochester, NY 14627, USA}
\affiliation{$^{2}$Department of Electrical and Computer Engineering, University of Rochester, Rochester, NY 14627, USA}

\begin{abstract}
This document provides supplementary information to “Chip-scale simulations in a quantum-correlated synthetic space”. It contains theoretical calculations to support some results in the main article and details on the device design and fabrication, and the experimental setup.
\end{abstract}

\maketitle

\onecolumngrid


\section{\label{sec:level1} Theoretical Model}
Theoretical work on a system of coupled cavity modes due to a time dependent modulation has been considered on several occasions, for example [1-3]. Here we will summarize the calculation of the Hamiltonian of this system. We will then tailor it to our device and solve the time evolution of the generated photon pairs.

\par Consider an optical field $\vec{E}$ in a medium with a time and space dependent susceptibility $\epsilon(r,t)$. The Hamiltonian can be written as
\begin{align}
   H=\int \vec{D} \cdot \vec{E} dV = \int \epsilon(r,t) \vec{E}  \cdot \vec{E} dV.
\end{align}
Setting electric field to be real  $\vec{E}=\frac{1}{2}(E+E^*)$, assuming all modes with same polarization, dropping vector notation, ignoring $EE$ and $E^{*}E^{*}$ terms with rotating wave approximation, and quantizing the field as
\begin{align}
   E=\sum_n \hat{a}_n(t) E_n(r,\omega_n).
\end{align}
Here $\hat{a}_n(t)$ represent the cavity modes we are interested in with a spatial profile given by $E_n(r,\omega_n)$ and a frequency $\omega_n$. Separating the intrinsic and modulation components of the susceptibility as $\epsilon(e,t)=\epsilon(r)+\Delta \epsilon(r,t)$, the Hamiltonian becomes
\begin{align}
   H=\frac{1}{4}\int \sum_{m,n} \hat{a}_m\hat{a}^\dag_n E_m(r,\omega_m)E_n^*(r,\omega_n) (\epsilon(r)+\Delta\epsilon(r,t)) dV + H.C.
\end{align}
We will set the temporal and spatial control over the refractive index to be independent such that $\Delta\epsilon(r,t)=\Delta\epsilon(r)M(t)$, where $M(t)$ is the applied modulation signal. We will consider two cases: 1) when $m=n$
\begin{align}
   H=&\frac{1}{4}\sum_{n} \int \hat{a}_n\hat{a}^\dag_n E_n(r,\omega_n)E_n^*(r,\omega_n) (\epsilon(r)+\Delta\epsilon(r)M(t)) dV + H.C. \nonumber \\
   =&\frac{1}{4}\sum_{n} \hat{a}_n\hat{a}^\dag_n \left[ \int dV \epsilon(r) |E_n(r,\omega_n)|^2 + M(t)\int dV \Delta\epsilon(r)|E_n(r,\omega_n)|^2 \right] + H.C.
\end{align}
For a ring resonator $E_n(r,\omega_n)=R(r)e^{in\phi}$, where $\phi$ is the azimuthal coordinate and $n$ now becomes the azimuthal mode order. The coordinate $r$ now represents the transverse plane only. Furthermore due to the periodic cycling of the field inside the resonator, we can decompose the susceptibility modulation as a Fourier series as $\Delta\epsilon(r)=\sum_p f_p e^{ip\phi}$. Then
\begin{align}
   \int dV \Delta\epsilon(r)|E_n(r,\omega_n)|^2 = 0, ~~\text{if}~~ p \neq 0.
\end{align}
Normalize $\frac{1}{4}\int dV \epsilon(r) |E_n(r,\omega_n)|^2=\frac{\hbar \omega_n}{2}$, and
\begin{align}
   \frac{M(t)}{4}\int dV \Delta\epsilon(r)|E_n(r,\omega_n)|^2=\frac{M(t)}{4}\int dV f_0 |E_n(r,\omega_n)|^2 = \frac{\hbar\Delta\omega M(t)}{2}.
\end{align}
Then
\begin{align}
   H=\sum_n \left[ \frac{\hbar\omega_n}{2}(\hat{a}_n \hat{a}^\dag_n + \hat{a}^\dag_n \hat{a}_n) +\frac{\hbar\Delta\omega M(t)}{2}(\hat{a}_n \hat{a}^\dag_n + \hat{a}^\dag_n \hat{a}_n) \right].
\end{align}
Using $[\hat{a}_m,\hat{a}^\dag_n]=\delta_{mn}$,
\begin{align}
   H_0=\sum_n \hbar (\omega_n + \Delta\omega M(t))\left[\hat{a}^\dag_n \hat{a}_n +\frac{1}{2} \right].
\end{align}
This is the free field part of the Hamiltonian with $\Delta\omega=\frac{1}{2\hbar} \int dV f_0 |E_n(r,\omega_n)|^2 $ being the modulation on the center frequency of the resonator modes. Note that $\Delta\omega$ will be zero if $f_0=0$, i.e. if the spatial dependence of the refractive index modulation has zero mean value.\\

Now consider 2) $m\neq n$, Then 
\begin{align}
    &\int dV \epsilon(r) E_m(r,\omega_m) E^*_n(r,\omega_n), \nonumber \\
    =&\int \epsilon(r) R_m(r,z)R^*_n(r,z)e^{i(m-n)\phi}rdrd\phi dz.
\end{align}
Since $\epsilon(r)$ has no $\phi$ dependence, then since $m \neq n$
\begin{align}
    \int rdrdz \epsilon(r) R_m(r,z)R^*_n(r,z) \int d\phi e^{i(m-n)\phi} = 0.
\end{align}
Now consider
\begin{align}
    &\int dV \Delta\epsilon(r) E_m(r,\omega_m) E^*_n(r,\omega_n), \nonumber\\
    =& \int \sum_p f_p e^{ip\phi} R_m(r,z)R^*_n(r,z) e^{i(m-n)\phi} rdrd\phi dz, \nonumber \\
    =& \sum_p f_p \int R_m(r,z)R^*_n(r,z) e^{i(m-n+p)\phi} rdrd\phi dz.
\end{align}
This is only non-zero if $m-n+p=0$. Therefore,
\begin{align}
    =2\pi \int f_{n-m}R_m(r,z)R^*_n(r,z) rdrdz,
\end{align}
and we get
\begin{align}
    & \frac{M(t)}{4}\int dV \Delta\epsilon(r) E_m(r,\omega_m) E^*_n(r,\omega_n), \nonumber \\
    =& \frac{2\pi M(t) f_{n-m} }{4} \int R_m(r,z)R^*_n(r,z) rdrdz \nonumber \\
    =& \hbar G_{m,n} M(t),
\end{align}
where $G_{m,n}=\frac{2\pi f_{n-m} }{4\hbar} \int R_m(r,z)R^*_n(r,z) rdrdz$ is the coupling strength between the m-th and n-th mode. This implies that the exact shape of the modulated region of the resonator is not relevant. As long as the modulation is not uniform across the entirety of the resonator, some spatial frequency component will match the momentum difference between the cavity modes resulting in a non-zero coupling. This gives the interaction part of the Hamiltonian $H_I$
\begin{align}
    H_I=\sum_{m,n} \hbar G_{m,n} M(t) \left[ \hat{a}_m \hat{a}^\dag_n + \hat{a}^\dag_m \hat{a}_n \right], ~~~ for~~ m\neq n,
\end{align}
and $H=H_0+H_I$
\begin{align}
    H=\sum_n \hbar (\omega_n + \Delta\omega M(t))\left[\hat{a}^\dag_n \hat{a}_n +\frac{1}{2} \right]   + \sum_{m\neq n} \hbar G_{m,n} M(t) \left[ \hat{a}_m \hat{a}^\dag_n + \hat{a}^\dag_m \hat{a}_n \right].
\end{align}
Consider an experimentally implementable situation where a sinusoidal modulation is electrically applied to the material susceptibility, i.e. $M(t)=cos(\Omega t)$, then $H_I$ will vanish unless $\omega_m-\omega_n=\Omega$. And $\Delta\omega$ will vanish if $\Delta\epsilon(\phi)$ is symmetric about zero, i.e has a zero DC value.\\
\par For the next section we will set $\Delta\omega=0$ since this is a small non-resonant contribution to the coupled power to the adjacent mode. We will also set $M(t)=cos(\Omega t)$ with the cavity FSR is equal to $\omega_n-\omega_{n-1}=\Omega$. Since the coupling extends only to the next mode, we can set $m=n-1$. The Hamiltonian becomes
\begin{align}
    H=\sum_n \hbar\omega_n \hat{a}^\dag_n \hat{a}_n + \sum_n \hbar G_{n,n-1}cos(\Omega t)[\hat{a}^\dag_n \hat{a}_{n-1} + \text{h.c.}].
\end{align}
Since $G_{m,n}=G(m-n)$ and $G_{n,n-1}=G(1)=G^*(-1)$. We assume $G$ is real, then $G(1)=G(-1)=G$
\begin{align}
    H=\sum_n \hbar\omega_n \hat{a}^\dag_n \hat{a}_n + \sum_n \hbar G cos(\Omega t)[\hat{a}^\dag_n \hat{a}_{n-1} + \text{h.c.}].
\end{align}
Moving into a rotating frame with $\hat{a}_n=\hat{a}^{'}_n e^{-i\omega_n t}$ and ignoring prime notation
\begin{align}
    H=\sum_n \frac{\hbar G}{2} (e^{i\Omega t}+e^{-i\Omega t})[\hat{a}^\dag_n \hat{a}_{n-1}e^{i\Omega t} + \text{h.c.}].
\end{align}
Applying rotating wave approximation
\begin{align}
    H_{RWA}=\sum_n \frac{\hbar G}{2} \hat{a}^\dag_n \hat{a}_{n-1} + \text{h.c.}
\end{align}
This is the tight-binding model of a network of modes and often appears when modelling crystal lattices. The rotating wave approximation also shows that if $\Omega$ is detuned from the mode spacing, $H_1$ will vanish which justifies our choice of setting $m=n-1$ in Eq. (16).

\par We can now modify this system to our waveguide-coupled resonator. Consider $m$ number of signal and idler modes in the frequency comb defined as $\hat{a}_{S,n}$, $\hat{a}_{I,n}$ with $n=1,2,3,...,m$, generated by a monochromatic pump $\hat{a}_p$. All operators are in a rotating frame of their resonance frequency as stated earlier i.e. $\omega_n$ and $\omega_{-n}$ for $\hat{a}_{S,n}$ and $\hat{a}_{I,n}$, respectively , respectively. The frequency comb structure is such that $\omega_n=\omega_0 +n\Omega_{FSR}$, where $\Omega_{FSR}$ is the dispersion-free resonator FSR. We will assume that the signal modes are separated enough from the idler modes such that their is no transfer  of photons between them, i.e. the photons considered here don't cross the spectral center. The interaction Hamiltonian for SPDC in a semi-classical treatment  can be written as 
\begin{align}
    H_{SPDC}= \sum_n \hbar g^{'} \hat{a}_p \hat{a}^{\dag}_{S,n} \hat{a}^{\dag}_{I,n} + H.C. = \sum_n \hbar g \hat{a}^{\dag}_{S,n} \hat{a}^{\dag}_{I,n} + \text{h.c.},
\end{align}
where $g^{'}$ is the second-order nonlinear coupling strength and we have absorbed the pump field into an effective nonlinear coupling strength $g$, treating it as a classical field. We also assume that the frequencies of pump, signal and idler photons match the SPDC process exactly as $\omega_p=\omega_n+\omega_{-n}$ which is true as long as $\omega_p =2\omega_0$ and the resonator's FSR is dispersionless. These criteria are met in our experiment. \\

\begin{figure}
\centering
  \includegraphics[scale=0.92]{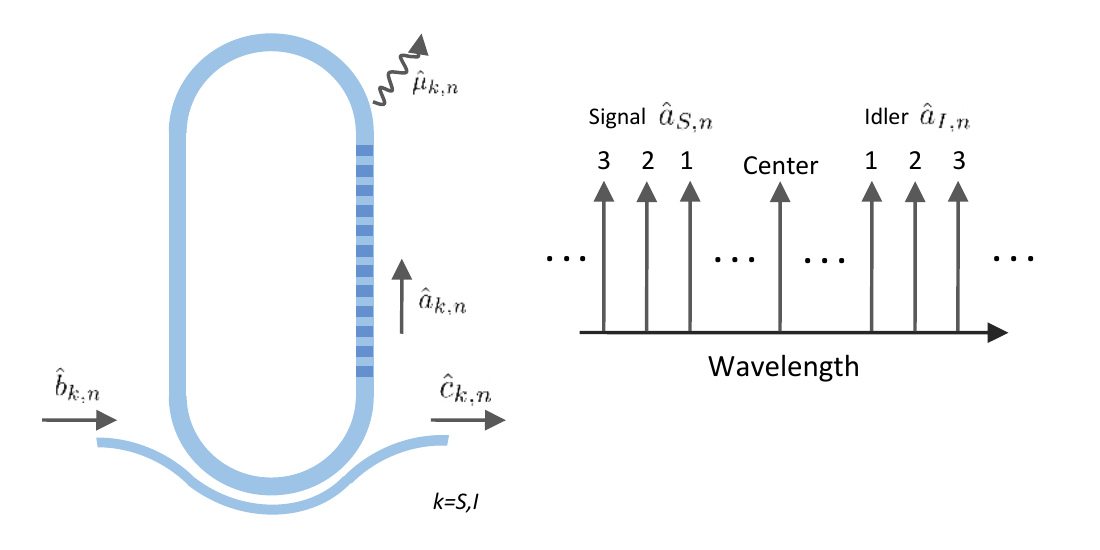}
  \caption{This figure is to support the theoretical treatment. The field operators in each section of the device are indicated. The modes of the generated frequency comb as used in the calculation are also indicated.}
  \label{figS1}
\end{figure}

\par
\begin{figure}
\centering
  \includegraphics[scale=0.95]{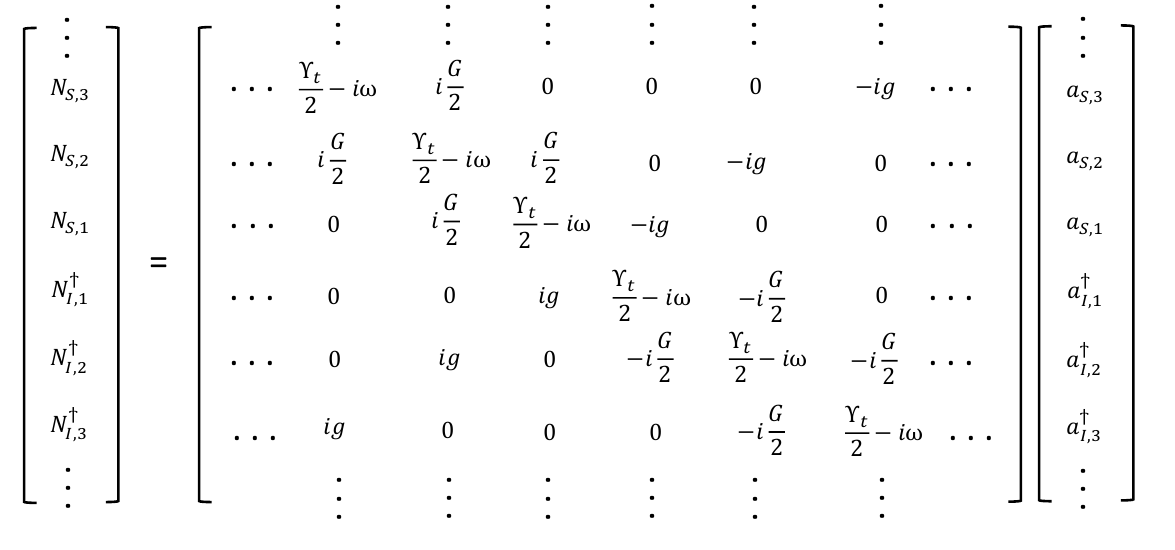}
  \caption{Matrix representation of Eqs. (26) and (27).}
  \label{figS2}
\end{figure}

 We will now introduce intrinsic loss of the resonator $\gamma_0$ and external coupling rate of the waveguide $\gamma_{ex}$. Finally, we can use input output relations of optical cavities and introduce three additional set of operators as
\begin{align}
    \hat{c}_{k,n}=\hat{b}_{k,n}+i\sqrt{\gamma_{ex}}\hat{a}_{k,n}, ~~~ \text{where}~ k=S,I,
\end{align}
where $\hat{b}_{k,n}$ is the noise operator for the field in the coupling waveguide corresponding to $\gamma_{ex}$ and $\hat{c}_{k,n}$ is the transmitted field operator as shown in Fig.~\ref{figS1}. We also introduce noise operator $\hat{\mu}_{k,n}$ corresponding to the resonator's intrinsic loss $\gamma_0$. We can now write a  modified Hamiltonian for a nonlinear coupled-mode resonator with photon pair generation as
\begin{align}
    H=H_{RWA}+H_{SPDC}=\sum_n \frac{\hbar G}{2} [\hat{a}^\dag_{S,n} \hat{a}_{S,n-1}+\hat{a}^\dag_{I,n} \hat{a}_{I,n-1}]+\sum_n \hbar g \hat{a}^{\dag}_{S,n} \hat{a}^{\dag}_{I,n}+ \text{h.c.}
\end{align}
The Heisenberg equation of motion for this Hamiltonian, accounting for losses and external coupling can be calculated to be
\begin{align}
    \frac{d\hat{a}_{k,n}}{dt}=\frac{1}{i\hbar}[\hat{a}_{k,n}, H]-\frac{\gamma_0+\gamma_{ex}}{2} \hat{a}_{k,n} +\sqrt{\gamma_0}\hat{\mu}_{k,n}+i\sqrt{\gamma_{ex}}\hat{b}_{k,n}.
\end{align}
The signal and idler equations then become
\begin{align}
    \frac{d\hat{a}_{S,n}}{dt}=-\frac{\gamma_t}{2}\hat{a}_{S,n} -i\frac{G}{2}[\hat{a}_{S,n-1}+\hat{a}_{S,n+1}]-ig\hat{a}^{\dag}_{I,n}+\hat{N}_{S,n},\\
    \frac{d\hat{a}_{I,n}}{dt}=-\frac{\gamma_t}{2}\hat{a}_{I,n} -i\frac{G}{2}[\hat{a}_{I,n-1}+\hat{a}_{I,n+1}]-ig\hat{a}^{\dag}_{S,n}+\hat{N}_{I,n},
\end{align}
where we have defined $\gamma_t=\gamma_0+\gamma_{ex}$ and $\hat{N}_{k,n}=\sqrt{\gamma_0}\hat{\mu}_{k,n}+i\sqrt{\gamma_{ex}}\hat{b}_{k,n}$. Taking Fourier transform of this system of equations, we obtain a set of 2$m$ linear equations in the frequency domain
\begin{align}
    \hat{N}_{S,n}(\omega)&=[\frac{\gamma_t}{2}-i\omega]\hat{a}_{S,n}(\omega)+i\frac{G}{2}[\hat{a}_{S,n-1}(\omega)+\hat{a}_{S,n+1}(\omega)]+ig\hat{a}^{\dag}_{I,n}(-\omega)\\
    \hat{N}^{\dag}_{I,n}(-\omega)&=[\frac{\gamma_t}{2}-i\omega]\hat{a}^{\dag}_{I,n}(-\omega)-i\frac{G}{2}[\hat{a}^{\dag}_{I,n-1}(-\omega)+\hat{a}^{\dag}_{I,n+1}(-\omega)]-ig\hat{a}_{S,n}(\omega)
\end{align}
This can be represented as a matrix system $\Bar{N}(\omega)=M(\omega) A(\omega)$ where $\Bar{N}(\omega)$ is the vector of noise operators $N_{k,n}$, $A(\omega)$ is the vector of signal and idler operators and $M(\omega)$ is the 2$m\times$2$m$ matrix of the system. Figure~\ref{figS2} shows the matrix system in detail. Taking the inverse of this matrix gives the solution for the intra-cavity signal and idler fields as

\begin{align}
    &\hat{a}_{S,n}(\omega)=\sum_{p=1:m} U_{m-n+1,p}(\omega) \hat{N}_{S,m-p+1}(\omega) + \sum_{q=m+1:2m} U_{m-n+1,q}(\omega) \hat{N}^{\dag}_{I,q-m}(-\omega) \\
    &\hat{a}_{I,n}^{\dag}(-\omega)=\sum_{p=1:m} U_{m+n,p}(\omega) \hat{N}_{S,m-p+1}(\omega) + \sum_{q=m+1:2m} U_{m+n,q}(\omega) \hat{N}^{\dag}_{I,q-m}(-\omega)
\end{align}

where $U=M^{-1}$ and is evaluated numerically with $U_{X,Y}$ being its elements. The second-order cross-correlation function $g^{(2)}_{X,Y}(\tau)$ for a mode pair $X,Y$ is given by (ignoring normalization)
\begin{align}
    g^{(2)}_{X,Y}(\tau)&=\langle \hat{c}^{\dag}_{S,X}(t)\hat{c}^{\dag}_{I,Y}(t+\tau)\hat{c}_{I,Y}(t+\tau)\hat{c}_{S,X}(t)\rangle, ~~~~ X,Y=1,2,3,...,m \nonumber \\ 
    &=\int\int\int\int d\omega_{1}d\omega_{2}d\omega_{3}d\omega_{4}\langle \hat{c}^{\dag}_{S,X}(\omega_1)\hat{c}^{\dag}_{I,Y}(\omega_2)\hat{c}_{I,Y}(\omega_3)\hat{c}_{S,X}(\omega_4)\rangle. e^{i[\omega_{1}t+\omega_{2}(t+\tau)-\omega_{3}(t+\tau)-\omega_{4}t]}
\end{align}
This gives us the joint spectral intensity (JSI) of the generated photons for each pair of modes. Evaluating this using Eqs. (21), (28) and (29), we get
\begin{align}
    \text{JSI}(\omega_{1},\omega_{2},\omega_{3},\omega_{4})=&4\pi^2\gamma^{2}_{ex} \left[ U^{*}_{m-X+1,m+y}(\omega_1) U_{m-X+1,m+y}(\omega_4) \right. \nonumber\\ 
    &\left. -\gamma_{t}U^{*}_{m-X+1,m+Y}(\omega_1) \sum_{p=m+1:2m} U^{*}_{m+Y,p}(-\omega_3)U_{m-X+1,q}(\omega_4) \right. \nonumber \\
    &\left. -\gamma_{t}U_{m-X+1,m+Y}(\omega_4) \sum_{q=m+1:2m} U^{*}_{m-X+1,q}(\omega_1)U_{m+Y,q}(-\omega_2) \right. \nonumber \\
    &\left.  +\gamma^{2}_{t} \sum_{p,q=m+1:2m} U^{*}_{m-X+1,q}(\omega_1)U_{m+Y,q}(-\omega_2)U^{*}_{m+Y,p}(-\omega_3)U_{m-X+1,p}(\omega_4) \right] \nonumber \\ &\times\delta(\omega_1+\omega_2)\delta(\omega_3+\omega_4).
\end{align}
\begin{figure}[h!]
\centering
  \includegraphics[scale=1]{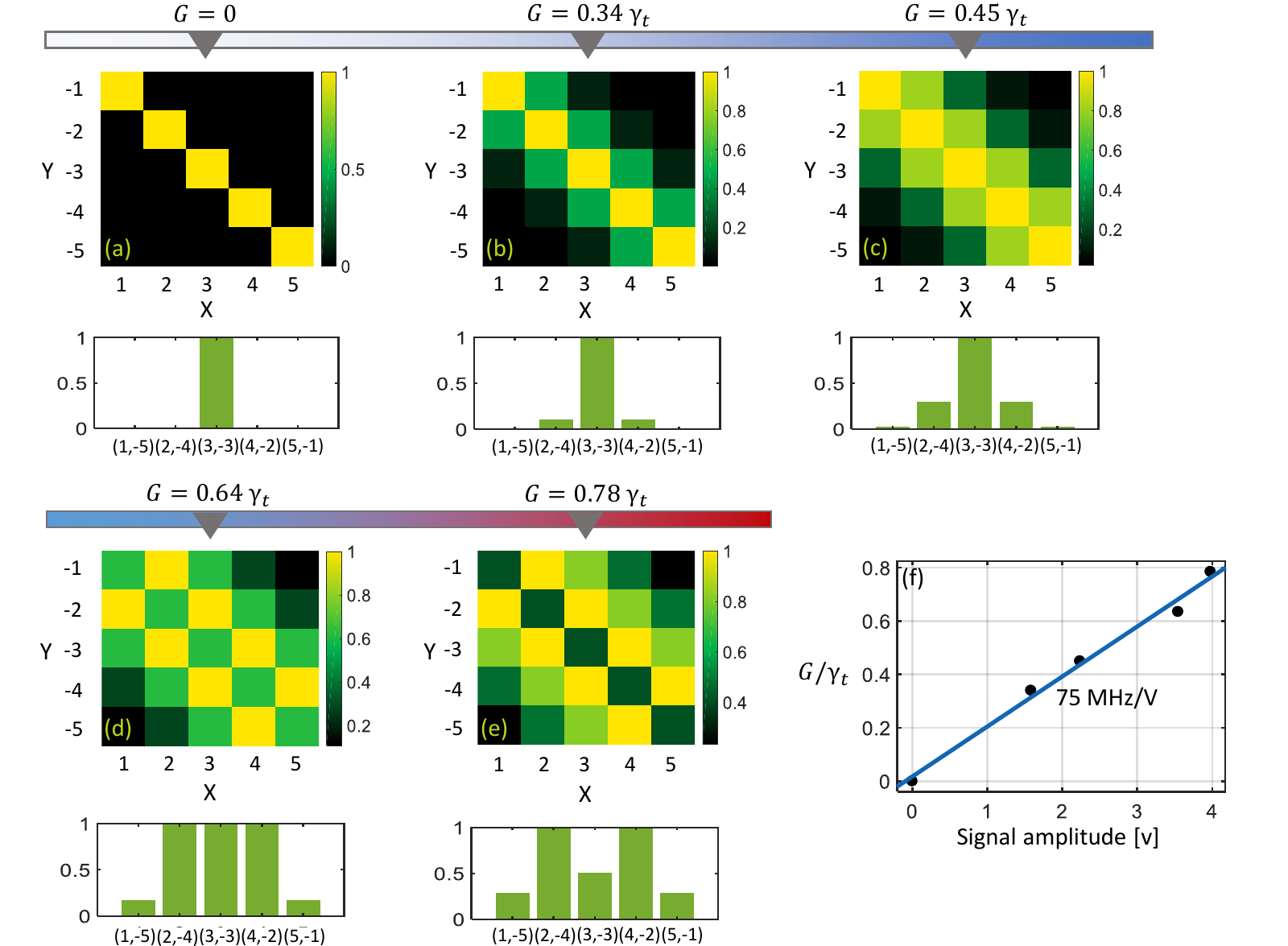}
  \caption{(a)-(e) Theoretically calculated bi-photon random walk plots using Eq. (31) to support the experimental data in Fig. 3 (main article). We have chosen five values of the coupling strength $G$ where the diagonal spread of the JSI matches with the experiment. Using these values we have obtained electro-optically-induced mode coupling efficiency as a linear fit of $G$ with the applied microwave signal voltage as shown in (f).}
  \label{figS3}
\end{figure}
Here, we have made use of the commutation relationships $[\hat{N}_{k,m}(\omega_{1}),\hat{N}^{\dag}_{l,n}(\omega_{2})]=2\pi \gamma_{t}\delta_{kl}\delta_{mn}\delta(\omega_{1}-\omega_{2})$, and $[\hat{b}_{k,m}(\omega_{1}),\hat{b}^{\dag}_{l,n}(\omega_{2})]=2\pi \gamma_{ex}\delta_{kl}\delta_{mn}\delta(\omega_{1}-\omega_{2})$. For simplicity, we set $\omega_1,..,\omega_4$ to zero to get the amplitude at the center of each resonance. This gauges the strength of the correlation at each mode pair and is equivalent to an integrated temporal correlation as was done for the data in Fig. 3 (main article). Figure~\ref{figS3} shows the results of this calculation for the resonator parameters we have measured. We observe good agreement with the experimental results obtained in Fig. 3 in the main article. By comparing the theory, we also obtain a maximum electro-optic modulation induced mode coupling amplitude of about $0.8\gamma_t$ (2$\pi\times$320 MHz). For the highest amplitude, we observe some deviation from the experimental data. The (1,-5) and (5,-1) terms have a much weaker amplitude in Fig.~\ref{figS3}(e) compared to the measurements in Fig. 3(f) in the main article. This is likely due to two contributing factors. First, the modulation of the pump mode may be causing off-diagonal photon pair generation. In our treatment we have ignored the random walk of the pump laser since the resonator FSR is 48 GHz around the pump wavelength compared to the 50 GHz microwave signal frequency. This suppresses any coupling of the pump to its adjacent modes. However at very strong modulation amplitudes some power tends to couple and generate photons off the main JSI diagonal. In that case the second order scattering terms such as (1,-5) and (5,-1) will get a small contribution from the adjacent pump modes. Second, at the highest amplitude used in the experiment, we are pushing the microwave amplifier towards nonlinearity. This can cause generation for higher harmonics of the signal at 100 GHz and 150 GHz. The second-harmonic can directly couple modes separated by two FSRs which can have a small contribution to these terms as well.

\subsection{Bloch Oscillations}
To theoretically model the two dimensional temporal oscillations of the bi-photon correlation function, it is more suitable to perform a closed system analysis of the resonator, setting loss and external coupling to zero and evaluating the joint spectrum of the photons inside the resonator as a function of time. This can be done by calculating the time evolution of the biphoton state generated by SPDC under the influence of the modulation Hamiltonian $H_{RWA}$ alone. We start by modifying The system to account for a detuning $\Delta$ in the modulation frequency given by $\Delta=(\omega_{n}-\omega_{n-1})-\Omega$. Due to this detuning, Eq. (19) becomes
\begin{align}
    H_{RWA}=\sum_n \frac{\hbar G}{2}e^{i\Delta t} \hat{a}^\dag_n \hat{a}_{n-1} + \text{h.c.}
\end{align}
Comparing this Hamiltonian to a solid state Hamiltonian under an action of an external electric field [4]
\begin{align}
    H_{SS}=\sum_{r_{1},r_{2}} C_{r_{1},r_{2}}e^{i\phi} \hat{b}^\dag_{r_1} \hat{b}_{r_2} + \text{h.c.},
\end{align}
where $C_{r_{1},r_{2}}$ is the hopping coefficient between lattice sits at positions $r_{1}$ and $r_{2}$, and $\hat{b}_{r_{n}}$ is the atomic operator for the localized electronic state at $r_{n}$. Here the phase $\phi$ of the hopping coefficient appears when an external gauge potential is applied given by
\begin{align}
    \phi=\frac{e}{\hbar} \int^{r_2}_{r_1} \Bar{A}\cdot d\Bar{l},
\end{align}
where $\Bar{A}$ is the external gauge potential. Comparing this to Eq. (32), we obtain a synthetic electric field for photons as
\begin{align}
    \int^{\omega_n}_{\omega_{n-1}} \Bar{A}\cdot d\omega=\Delta t, ~~~ E=\frac{dA}{dt}=\frac{\Delta}{\text{FSR}}.
\end{align}
This shows how a detuned microwave drive can simulate a constant electric in a crystal lattice which leads to Bloch oscillations of the electron's wavefunction.
\par
\begin{figure}[h!]
\centering
  \includegraphics[scale=1]{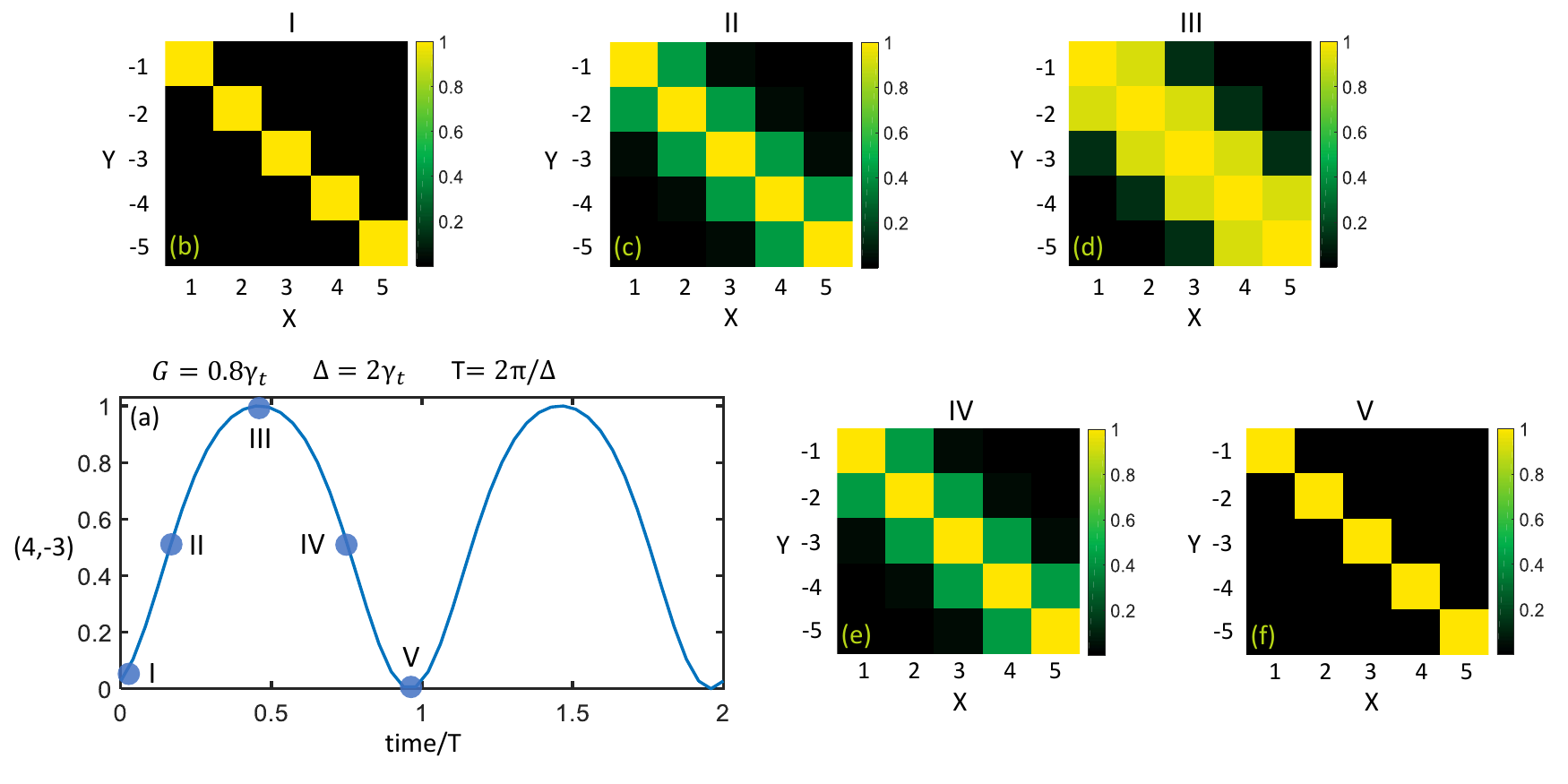}
  \caption{Bloch oscillations of the bi-photon correlation function inside the resonator plotted using Eq. (40) for a modulation strength $G=0.8\gamma_{t}$ and a detuning $\Delta=2\gamma_{t}$. In (a), we have plotted the JSI strength at (4,-3) during the time evolution of the correlation function with the matrices in (b)-(f) plotting the entire JSI at five points in time as indicated in the plot in (a).}
  \label{figS4}
\end{figure}

 Going back to Eq. (32), we can get a system of equations similar to Eqs. (23-25) as 
\begin{align}
    \frac{d\hat{a}_n}{dt}=&\frac{1}{i\hbar}[\hat{a}_n, H_{RWA}] \nonumber \\
    =&-i\frac{G}{2}e^{i\Delta t}\hat{a}_{n-1}-i\frac{G}{2}e^{-i\Delta t}\hat{a}_{n+1}.
\end{align}
This is a system of linearly-coupled differential equations and can be represented as $X^{'}(t)=M(t)X(t)$, where $X$ is the vector of the field operators $\hat{a}_n$. We can solve an initial value problem setting the initial state to be an equal probability superposition state of $m$ modes as
\begin{align}
    \Psi(t=0)=\frac{1}{N}\sum_{n=1:m}\hat{a}^{\dag}_{n}\hat{a}^{\dag}_{-n}| 0\rangle 
\end{align}
The state at a later time $t$ can obtained as 
\begin{align}
    X(t)= e^{\int^{t}_0 M(t) dt}X(0),
\end{align}
where $X(0)$ contains coefficients of the initial state in Eq. (37). The state at a later time is given
\begin{align}
    \Psi(t)=\frac{1}{N}\sum_{m,n,p=1:m} U^{* ~-1}_{-n,m}U^{* ~-1}_{n,p}   \hat{a}^{\dag}_{m}\hat{a}^{\dag}_{p}| 0\rangle,
\end{align}
where $U=e^{\int^{t}_0 M(t) dt}$ and $U^{* ~-1}$ is the complex conjugate of the inverse of the matrix $U$ with $U_{m,n}$ being its elements. The joint spectral intensity (JSI) for a mode pair ($X,Y$) is given by the modulus squared of $\Psi(t)$ at that point given by
\begin{align}
    \text{JSI}(X,Y)=\left|\frac{1}{N}\sum_{n=1:m} U^{* ~-1}_{-n,X}U^{* ~-1}_{n,Y} \right|^{2}.
\end{align}
Figure~\ref{figS4} shows the evolution of the JSI for a detuning $\Delta=2\gamma_{t}$ and a modulation amplitude $G=0.8 \gamma_{t}$. Here we have plotted the JSI strength at a point one mode offset from the main correlation diagonal i.e. at $(X+1,-X+1)$, as was done in the experiment. Any power at this point is due to the photons jumping into this mode due to the modulation. Figure~\ref{figS4} shows that the JSI has an oscillatory evolution at a frequency matching $\Delta$, verifying a simulation of Bloch oscillations.

\par
\begin{figure}[h!]
\centering
  \includegraphics[scale=1]{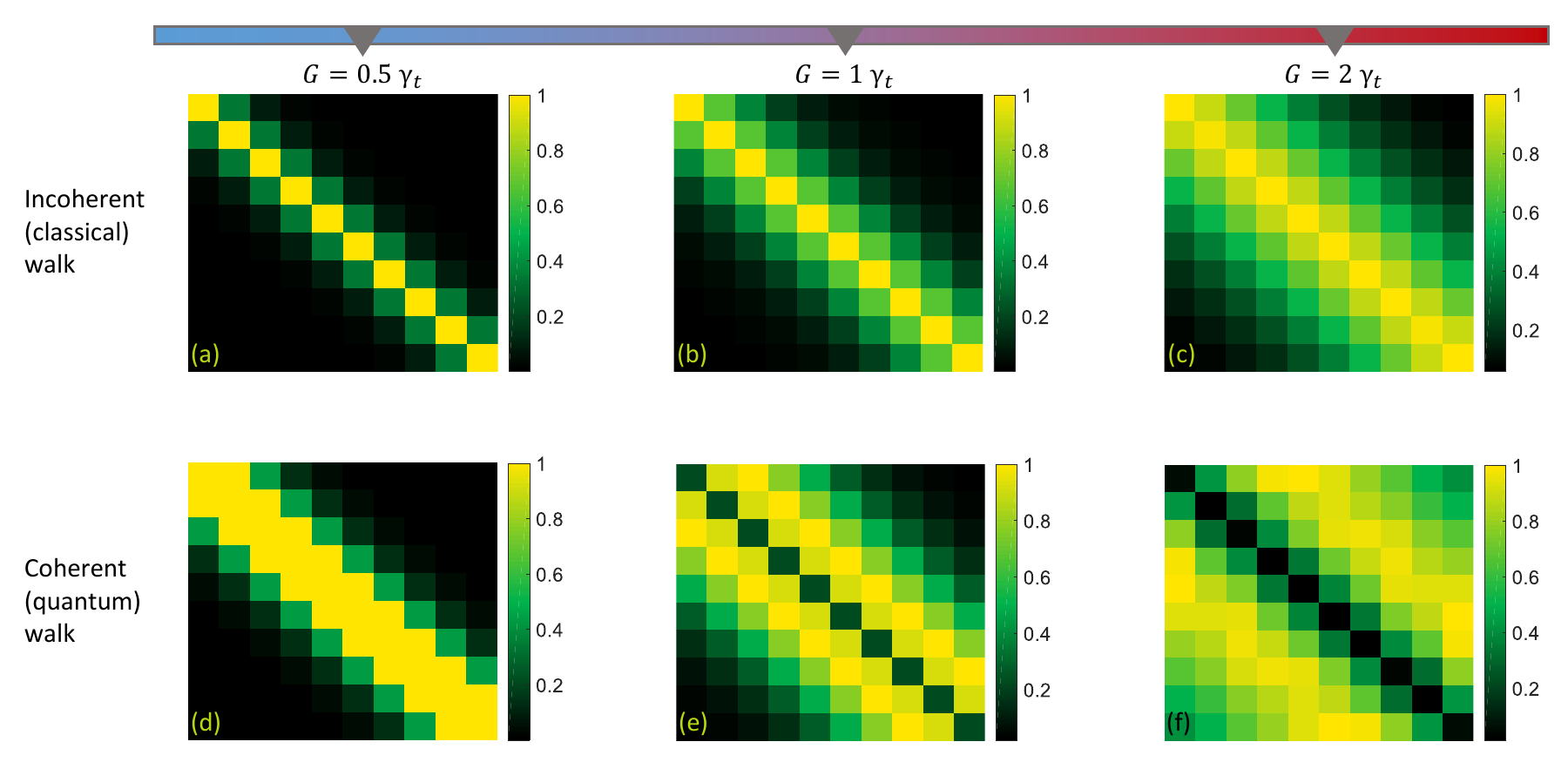}
  \caption{Comparison of random walks of an incoherent mixture (a)-(c), and a coherent superposition (d)-(f) of frequency modes generated by SPDC with increasing modulation amplitude indicated at the top.}
  \label{figS7}
\end{figure}

 Random walk of quantum correlated photons manifest features that are not present if classical states of light are used (for example see [5]). In Fig.~\ref{figS7} we theoretically compare the random walk of quantum-correlated photons with a incoherent mixture of the same frequency correlations. This is done by incoherently adding the random walk of each term of the superposition $\sum_{n} \hat{a}^{\dag}_{n}\hat{a}^{\dag}_{-n}| 0\rangle $ to simulate a statistical mixture of frequency correlated modes [6]. As we see in Fig.~\ref{figS7}, the two walks are quite different at high modulation amplitudes. In the incoherent walk, the correlations do not decrease in the main diagonal, even at high modulation amplitudes and the photons tend to confine near the center. On the other hand, in the coherent superposition, the photons spread laterally very quickly, and the correlations in the main diagonal drop below the correlations off diagonal, as was seen in the experimental data in Fig. 4 (main article). Here we observe the tendency of the photons to bunch together, i.e. both photons tend to move towards the blue side or the red side of the main diagonal, together. This is a frequency analog of the Hong-Ou-Mandel interference in this higher dimensional space. This also demonstrates the coherent nature of the frequency superposition state that is generated in the frequency comb.
 
\begin{figure}[h!]
\centering
  \includegraphics[scale=1]{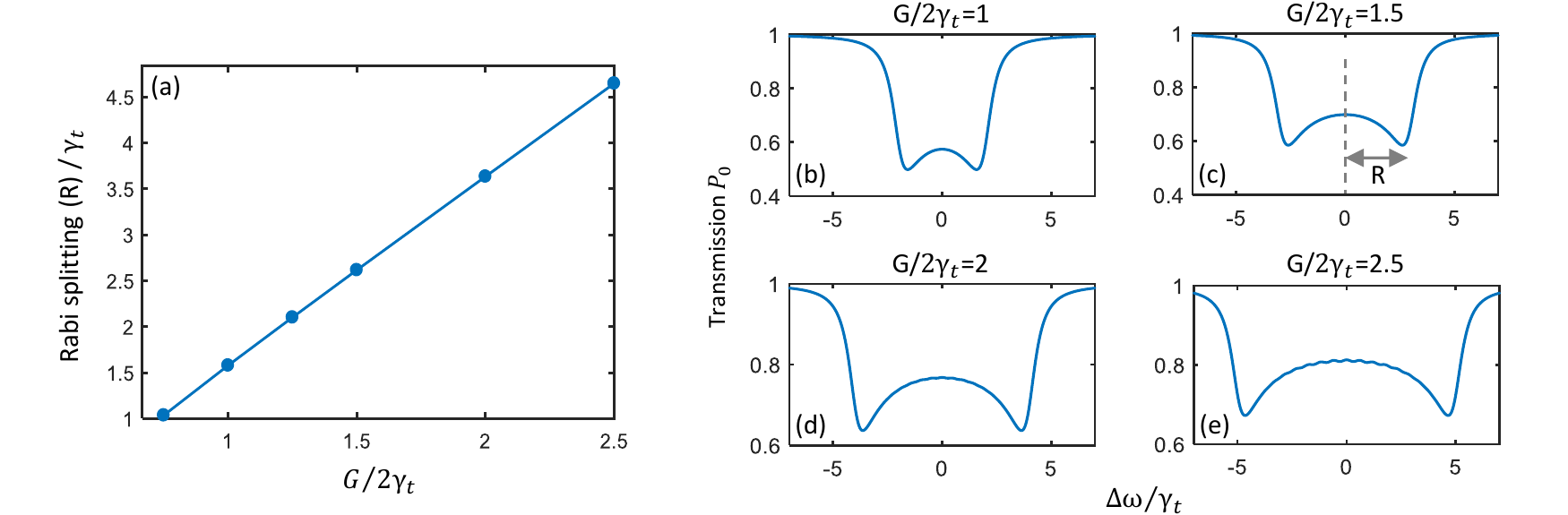}
  \caption{(a) Rabi splitting (R) of the resonances at different modulation amplitudes. (b)-(e) Laser transmission plotted against laser detuning $\Delta\omega$ for some of the points in (a) using Eq. (44).}
  \label{figS8}
\end{figure}

\subsection{Strong coupling regime and Rabi oscillations}
The theoretical treatment can be used to obtain the bi-photon temporal correlation function $g^{(2)}_{X,Y}(\tau)$ by taking a four-dimensional inverse Fourier transform of the JSI as stated in Eq. (30). This model works in the strong coupling regime as well. This is used to obtain theory plots for Rabi oscillations Fig. 5 in the main article. To fit the data, we need an accurate value for the coupling strength $G$. We do this by measuring the mode-splitting induced by strong coupling of adjacent resonances by scanning a laser through a resonance and measuring the average output power. Theoretically, we can model this by slightly modifying Eq. (24) as

\begin{align}
    \frac{da_{n}}{dt}=-\frac{\gamma_t}{2}a_{n} -i\frac{G}{2}[a_{n-1}+a_{n+1}]+i\sqrt{\gamma_{ex}}A_{in}e^{-i\Delta\omega t}\delta_{n,0}.
\end{align}
Here we have ignored the SPDC term and introduced a CW laser input with a detuning $\Delta\omega$ to the mode with the index 0. Going into a frame rotating with $\Delta\omega$, we get
\begin{align}
    \frac{da_{n}}{dt}=(i\Delta\omega-\frac{\gamma_t}{2})a_{n} -i\frac{G}{2}[a_{n-1}+a_{n+1}]+i\sqrt{\gamma_{ex}}A_{in}\delta_{n,0}.
\end{align}
We can get a steady-state solution by setting $\frac{da_{n}}{dt}=0$ and obtaining a system of linear equations.
\begin{align}
   a_{n}= \frac{\frac{G}{2}[a_{n-1}+a_{n+1}]-i\sqrt{\gamma_{ex}}A_{in}\delta_{n,0}}{i\Delta\omega-\gamma_{t}/2}.
\end{align}
The transmitted power at the laser frequency in the zeroth mode is give by
\begin{align}
   P_{0}=|A_{in}+i\sqrt{\gamma_{ex}}a_{0}|^{2}.
\end{align}
Figure \ref{figS8} plots the transmitted power with a laser scan over a resonance using Eq. (44). We clearly observe the resonance splitting into a pair of dressed modes. We also plot the mode splitting as a function of the coupling strength. For our multi-mode system, the Rabi splitting is nearly twice more than what is obtained in a two-level system. Using these resits and the experimentally obtained laser transmission trace plotted in Fig. 5 (main article), we obtain $G/2\gamma_{t}=1.9$, putting us in the strong coupling regime. This measurement was done with a different device which had an optical quality factor twice as high as the one used for the random walk and Bloch oscillations. More details on the device are given in section II.

\begin{figure}
\centering
  \includegraphics[scale=1]{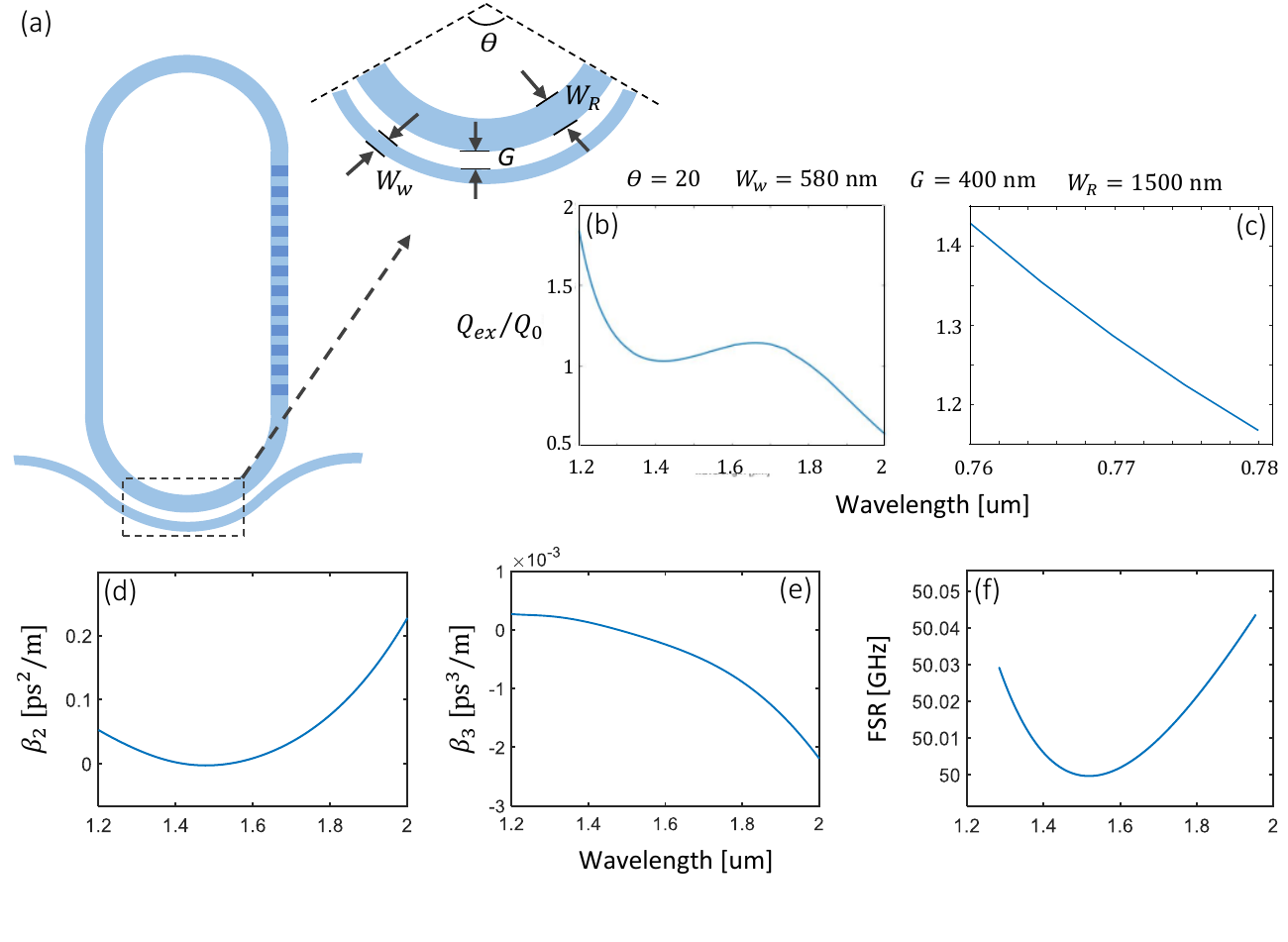}
  \caption{(a) Wrap-around geometry of the coupling waveguide with four design parameters: the waveguide width $W_w$, the coupling angle $\theta$, the waveguide ring gap $G$, and the ring width $W_R$. The ratio of the external quality factor $Q_{ex}$ to the intrinsic quality factor $Q_0$ is plotted for 1.2 $\mu$m - 2 $\mu$m wavelength region for the signal and idler photons in (b) and for pump wavelengths in the 0.76 $\mu$m - 0.78 $\mu$m wavelength region in (c). These are simulated with the optimized design parameters indicated above the plots. The resonator dispersion in the straight sections is calculated with the second-order dispersion coefficient $\beta_2$ plotted in (d), and the third-order dispersion coefficient $\beta_3$ plotted in (e). Using these parameters, the FSR of the resonator is plotted in (f) using a 50 GHz nominal value.}
  \label{figS5}
\end{figure}

\section{Device Design}
As stated in the main article, the device is designed to fulfill three requirements. First, the resonator width and etching depth are adjusted to set the zero dispersion wavelength of the fundamental TE mode at the center of the generated spectrum around 1550 nm. This simulation is done only on the straight section of the resonator which contributes about 80\% to the total length and thus the dispersion. This allows some deviation from zero due to the curved sections of the resonator where the refractive index is different due to the birefringence of the LN crystal. This is what we believe limits our SPDC bandwidth to about 40 THz. From the simulations, we obtain a resonator width of 1.5 $\mu$m and a 50\% etching depth of 300 nm. The length of the resonator from top to bottom is about 1.2 mm and is set to obtain a 50 GHz FSR. The simulated dispersion parameters are shown in Fig.~\ref{figS5}(d),(e) matching our criteria and the resulting FSR is simulated in Fig.~\ref{figS5}(f) using [7]. Here we see very little deviation in the FSR from a nominal value of 50 GHz indicating that both a broadband QOFC and a broadband electro-optic comb can be generated.

\par Second, we design a coupling waveguide to have a broadband dispersionless coupling to extract the entire  frequency comb. We utilize a wrap-around geometry for the waveguide. This gives three adjustable parameter to set: the coupling angle, gap and the waveguide width as shown in Fig.~\ref{figS5}(a). The forth parameter, i.e. the resonator width, has to be fixed based on the dispersion requirements as discussed earlier. We then solve a parameter optimization problem using these parameters to calculate the ring, waveguide coupling rate by evaluating the overlap of the two fields [8]. The optimization has two goals: to obtain broadband critical coupling in the telecom band, and to obtain as close to critical coupling as possible for the pump wavelengths in the 765 - 780 nm spectral region. Figure~\ref{figS5}(b),(c) show one of the obtained solutions which meets these two requirements. 

\par Finally, the two pairs of electrodes are designed. The poling period on the periodic electrodes is set to quasi phase-match second harmonic generation (SHG) for the fundamental TE modes at 1550 nm and 775 nm. This evaluates to 4.1 $\mu$m. This period is swept and we find that a period of 4.17 $\mu$m gives the strongest SH light which is then used for the SPDC experiment. The flat electrodes are designed with the impedance of the microwave mode at 50 GHz matching the 50 $\Omega$ impedance of the microwave signal generator to maximize power transfer. This gives a width of 50 $\mu$m. A gap of 2.5 $\mu$m from the sidewall of the resonator on each side gives decent coupling with the optical mode without degrading the optical quality factor too much. Nevertheless, we observe a 33\% reduction in the quality factor due the the two pairs of electrodes. 

\par
\begin{figure}
\centering
  \includegraphics[scale=0.9]{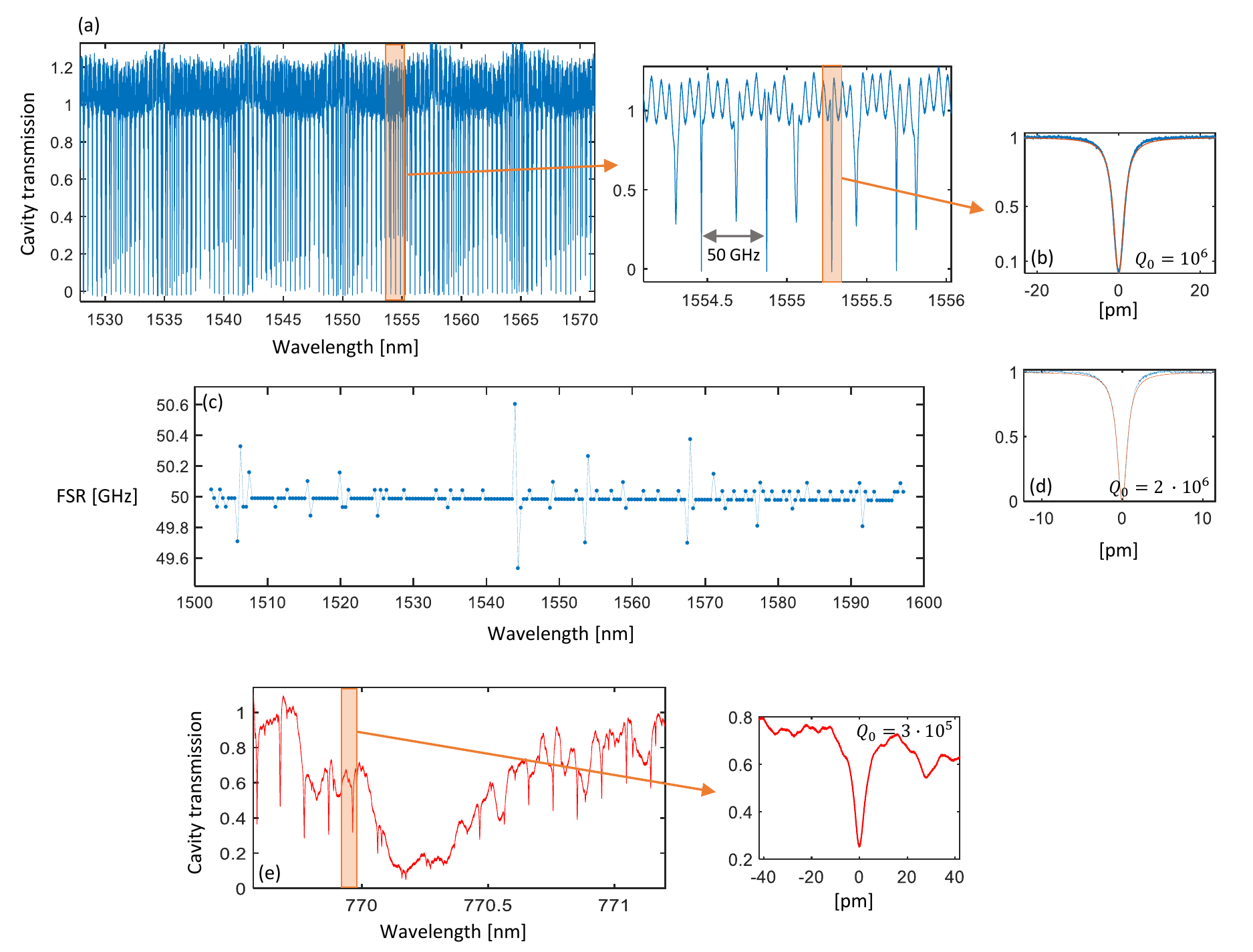}
  \caption{Device characterization. (a) Transmission spectrum of the lithium niobate resonator in the telecom band obtained by scanning a tunable laser through the coupling waveguide. The inset shows a part of the scan in more detail. (b) A detailed scan of one of the fundamental TE modes of the resonator that takes part in the SPDC interaction. The measured FSR of the resonator in this wavelength band is plotted in (c). (d) A scan of one of the fundamental TE modes of a different resonator that was annealed demonstrating a reduced linewidth. A similar wavelength scan in the near-infrared region is shown in (e) with the inset showing the resonance used for the pump in the SPDC process. }
  \label{fig2}
\end{figure}

 The device was fabricated on a 600 nm-thick x-cut LN thin film on a 4.7-$\mu$m silicon dioxide bottom cladding layer and a silicon substrate (NanoLN). The waveguide was first patterned with e-beam resist (ZEP-520A) via e-beam lithography, which was then transferred to the LN layer by 300 nm etching using Ar$^+$ ion milling. After resist removal, a second e-beam exposure is performed to pattern the electrode structures. The electrodes (20 nm Ti/400 nm Au) were then deposited via electron-gun evaporation, which is then followed by a standard lift-off process. After fabrication, the resonator was poled using a sequence of 240 V 10 ms square-wave electrical pulses applied to the periodically patterned electrodes. The poling efficiency is monitored heuristically by optimizing second harmonic generation (SHG). A tunable laser is scanned through the resonator while simultaneously applying the poling pulses. The pulses are applied until the SH signal power saturates. To further increase the quality factor for measurements in the strong coupling regime in Fig. 5 (main article), a fabricated device was annealed at 450$ ^\text{o}$ C for one hour in oxygen flow. This doubled the intrinsic quality factor as shown in Fig.~\ref{fig2}(d), reducing the loaded linewidth to 200 MHz.

\par The linear properties of the fabricated device are characterized by scanning a tunable laser through the coupling waveguide and measuring its transmission. The resonator transmission at telecommunication wavelengths where the photon pairs are generated is shown in Fig.~\ref{fig2}(a),(b), and at the pump wavelengths around 775 nm in Fig.~\ref{fig2}(e). Here, we can see that broadband critical coupling is achieved in the telecommunication region, while near-infrared modes are also accessible for pumping the SPDC process. We measure an intrinsic quality factor of $10^{6}$ for the telecom band modes and $3 \times 10^{5}$ for the near-infrared wavelength modes. Furthermore, the measured FSR of the resonator at the signal/idler wavelengths shows remarkable uniformity as shown in Fig.~\ref{fig2}(c).

\begin{figure}
\centering
  \includegraphics[scale=0.8]{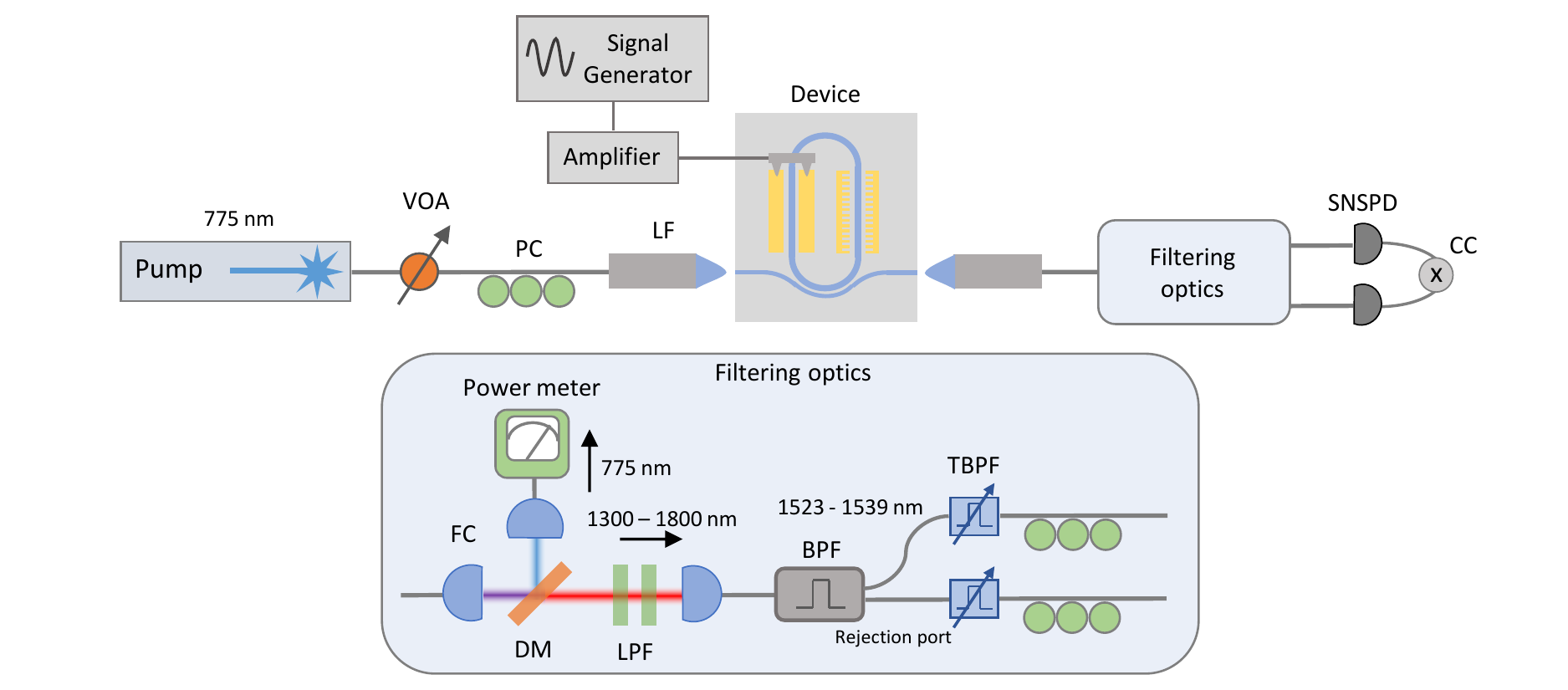}
  \caption{Illustration of the experimental setup. VOA: variable optical attenuator, PC polarization controller, LF: lensed fiber, SNSPD: superconducting nanowire single-photon detector, CC: coincidence counter, FC: fiber coupler, DM: dichroic mirror, LPF: low-pass filter, BPF: band-pass filter, TBPF: tunable band-pass filter.}
  \label{figS6}
\end{figure}

\section{Experimental setup}
The experimental setup is shown in Fig.~\ref{figS6}. Light is coupled in and out of the chip using a pair of lensed fibers. The chip is placed on a thermo-electric cooler (TEC) to control its temperature with feedback to maintain the resonance positions during the experiment. A tunable laser (NewFocus TLB-6712) drives the pair generation process. At the output end of the device, the laser and the generated photons are coupled out through the lensed fiber and passed through a sequence of filters. First, the laser is removed using a dichroic beam splitter and a sequence of two low-pass filters which together give an estimated pump extinction of 120 dB. Then the signal and idler photons are separated first using a bandpass filter which separates a 17-nm section of the spectrum centered at 1531 nm and sends the remaining spectrum through its rejection port. The two outputs of the bandpass filter are then sent to two tunable bandpass filters with a 3-dB bandwidth of 0.25 nm which can select individual modes of the resonator. The outputs of these filters are then sent to two superconducting nanowire single-photon detectors (SNSPDs) with their outputs fed to a coincidence counter for correlation measurements. The device is also electrically driven with a microwave signal generator at 50 GHz. the signal is first fed to an amplifier (Spacek Labs) with a 40 dB gain. The amplifier's output is then fed to the on-chip electrodes through a high frequency probe (FormFactor Infinity).
